\documentclass[review]{jfm}
\usepackage{amsmath,color,graphicx,amssymb}
\usepackage{natbib}
\usepackage{amsbsy}
\usepackage{latexsym}
\usepackage{subfigure}
\usepackage[normalem]{ulem}
\usepackage[english]{babel}
\usepackage{ifsym}
\usepackage{bbding}
\usepackage{wasysym}
\usepackage{epsfig} 
\usepackage{epsf,psfig}
\usepackage{epstopdf}
\usepackage{multirow}
\usepackage{array}
\usepackage{float}
\usepackage[english]{babel}
\usepackage{verbatim}
\usepackage{soul}
\usepackage[toc,page]{appendix}
\usepackage{setspace}
\usepackage{threeparttable}
\usepackage{booktabs,array}

\definecolor{gray}{rgb}{0.5,0.5,0.5}
\definecolor{black}{rgb}{0,0,0}
\definecolor{purple}{rgb}{0.57,0,0.86}
\definecolor{greenish}{rgb}{0.2,0.7,0.2}
\definecolor{orange}{rgb}{1,0.5,0}

\newcommand{\dbq}[1]{{{\color{blue} #1}}}
\newcommand{\kwm}[1]{{{\color{purple} #1}}}

\newcommand\mb{\mathbf}

\def\ks{k^*}
\def\As{A^*}

\def\etal{\mbox{\textit{et al.}}}
\def\beq{ \begin{equation}}
\def\eeq{\end{equation}}
\def\beqar{ \begin{eqnarray} }
\def\eeqar{ \end{eqnarray} }

\def\vK{von K\'{a}rm\'{a}n }

\begin{document}


\title{Inviscid Scaling Laws of a Self-Propelled Pitching Airfoil}

\author[K.W. Moored \etal]
{By K\ls E\ls I\ls T\ls H\ns W.\ns  M\ls O\ls O\ls R\ls E\ls D,$^1$ \ns
\and D\ls A\ls N\ls I\ls E\ls L\ns B.\ns Q\ls U\ls I\ls N\ls N,$^2$ \ns}

\affiliation{
$^1$Department of Mechanical Engineering and Mechanics \\
Lehigh University,
Bethlehem, PA 18015, USA\\
$^2$Department of Mechanical and Aerospace Engineering \\
University of Virginia,
Charlottesville, VA 22903, USA\\
[\affilskip]
}

\pubyear{2015}
\volume{}
\pagerange{}
\date{}

\maketitle


\begin{abstract}

Inviscid computational results are presented on a self-propelled virtual body combined with an airfoil undergoing pitch oscillations about its leading-edge.  The scaling trends of the time-averaged thrust forces are shown to be predicted accurately by Garrick's theory.  However, the scaling of the time-averaged power for finite amplitude motions is shown to deviate from the theory.  Novel time-averaged power scalings are presented that account for a contribution from added-mass forces, from the large-amplitude separating shear layer at the trailing-edge, and from the proximity of the trailing-edge vortex.  Scaling laws for the self-propelled speed, efficiency and cost of transport ($CoT$) are subsequently derived.  Using these scaling relations the self-propelled metrics can be predicted to within 5\% of their full-scale values by using parameters known \textit{a priori}.  The relations may be used to drastically speed-up the design phase of bio-inspired propulsion systems by offering a direct link between design parameters and the expected $CoT$.  The scaling relations also offer one of the first mechanistic rationales for the scaling of the energetics of self-propelled swimming.  Specifically, the cost of transport is shown to scale predominately with the added mass power.  This suggests that the $CoT$ of organisms or vehicles using unsteady propulsion will scale with their mass as $CoT \propto m^{-1/3}$, which is indeed shown to be consistent with existing biological data. 
\end{abstract}

\section{Introduction}
Over the past two decades, researchers have worked towards developing bio-inspired technologies that can match the efficiency, stealth and maneuverability observed in nature \cite[]{Anderson1998,Moored2011b,Moored2011,Curet2011,Ruiz2011,Jaworski2013,Gemmell2014}.  To this end, many researchers have detailed the complex flow physics that lead to efficient thrust production \cite[]{Buchholz2008,Borazjani2008b,Borazjani2009,Masoud2010,Dewey2011,Moored2012,Moored2014,Mackowski2015}.  Some have distilled these flow phenomena into scaling laws under fixed-velocity, net-thrust conditions \cite[]{Green2011,Kang_2011,Dewey_2013,Quinn2014b,Quinn2014,Das2016}.  A few others have developed scaling laws of self-propelled swimming to elucidate the important physical mechanisms behind free-swimming organisms and aid in the design of efficient self-propelled technologies. For instance, \cite{Bainbridge1957} was the first to propose that the speed of a swimming fish was proportional to its frequency of motion and tail beat amplitude.  This empirical scaling was given a mechanistic rationale by considering the thrust and drag scaling relations for animals swimming in an inertial Reynolds number regime \cite[]{Gazzola2014}.  Further detailed mathematical analyses have also revealed scaling relations for the self-propelled swimming speed of a flexible foil with variations in its length and flexural rigidity \cite[]{Alben_2012}.  However, these scaling relations only consider the speed and not the energetics of self-propelled swimming.

There is a wide variation of locomotion strategies used in nature that can modify the flow physics and energetics of a self-propelled swimmer. A large class of aquatic animals propel themselves by passing a traveling wave down their bodies and/or caudal fins.  These animals can be classified depending upon the wavelength, $\lambda$, of the traveling wave motion compared to their body length, $L$.   Consequently,  they fall along the undulation-to-oscillation continuum \cite[]{Sfakiotakis1999} where undulatory aguilliform swimmers such as eel and lamprey, have low non-dimensional wavelengths ($\lambda/L < 1$) and their entire body generates thrust.  On the oscillatory end of the continuum, subcarangiform, carangiform and thunniform swimmers such as trout, mackerel and tuna, respectively, have high non-dimensional wavelengths ($\lambda/L > 1$) and they also have distinct caudal fins that generate a majority of their thrust \cite[]{Lauder2006}.  In numerous studies, the caudal fins of these swimmers have been idealized as heaving and pitching wings or airfoils.  

Following this idea, classical unsteady wing theory \cite[]{Wagner1925,Theodorsen1935,Garrick1936,VonKarman1938} has been extended to analyze the performance of caudal fins in isolation; especially those with high aspect ratios \cite[]{Chopra1974,Chopra1976a,Chopra1977,Cheng1984a,Karpouzian1990a}.  When coupled with a drag law these theories can provide scaling relations for the swimming speed and energetics of self-propelled locomotion.  In fact, both potential flow-based boundary element solutions \cite[]{Jones1997,Quinn2014} and experiments \cite[]{Dewey_2013,Quinn2014,Mackowski2015} have shown that Garrick's theory \cite[]{Garrick1936} accurately captures the scaling trends for the thrust production of pitching airfoils.  However, it has recently been appreciated that the predicted Garrick scaling relation for the power consumption of a finite-amplitude pitching airfoil in a \textit{fixed velocity flow} is not in agreement with the scaling determined from experiments and boundary element computations \cite[]{Dewey_2013,Quinn2014}.  

Motivated by these observations, we extend previous research by considering three questions:  (1) How well does Garrick's linear theory predict the performance and scaling laws of \textit{self-propelled} swimmers, (2) going beyond linear theory what are the missing nonlinear physics that are needed to account for finite amplitude pitching motions, and (3) based on these nonlinear physics what are the scaling laws for self-propelled swimmers that use finite amplitude motions and are these laws reflected in nature?  By answering these questions, this work advances previous studies in a variety of directions.  First, the self-propelled performance of a virtual body combined with an airfoil pitching about its leading edge and operating in an inviscid environment is determined.  This combined body and `fin' produces a large parametric space that includes body parameters such as its wetted area, drag coefficient, relevant drag law and mass.  Second, novel physical insights are used to introduce new thrust and power scaling relations that go beyond previous scaling arguments \cite[]{Garrick1936,Dewey_2013,Quinn2014}.  Third, scaling relations for the speed and energetic performance of a self-propelled swimmer are subsequently developed.  These relations can predict speed, efficiency and cost of transport to within 5\% of their full-scale values by using parameters that are known \textit{a priori} after tuning a handful of coefficients with a few simulations. Lastly, the self-propelled scaling relations give the first mechanistic rationale for the scaling of the energetics of a self-propelled swimmer and they further generalize the swimming speed scaling relations developed by \cite{Gazzola2014}.  Specifically, the cost of transport of an organism or device using unsteady locomotion is proposed to scale predominately with the added mass power and consequently with its body mass as $CoT \propto m^{-1/3}$.  This scaling law is further shown to be consistent with existing biological data.


\section{Problem Formulation and Methods}
\subsection{Idealized Swimmer}
\begin{figure}
	\begin{center}
	     \centerline{
		\includegraphics[width=0.99\textwidth]{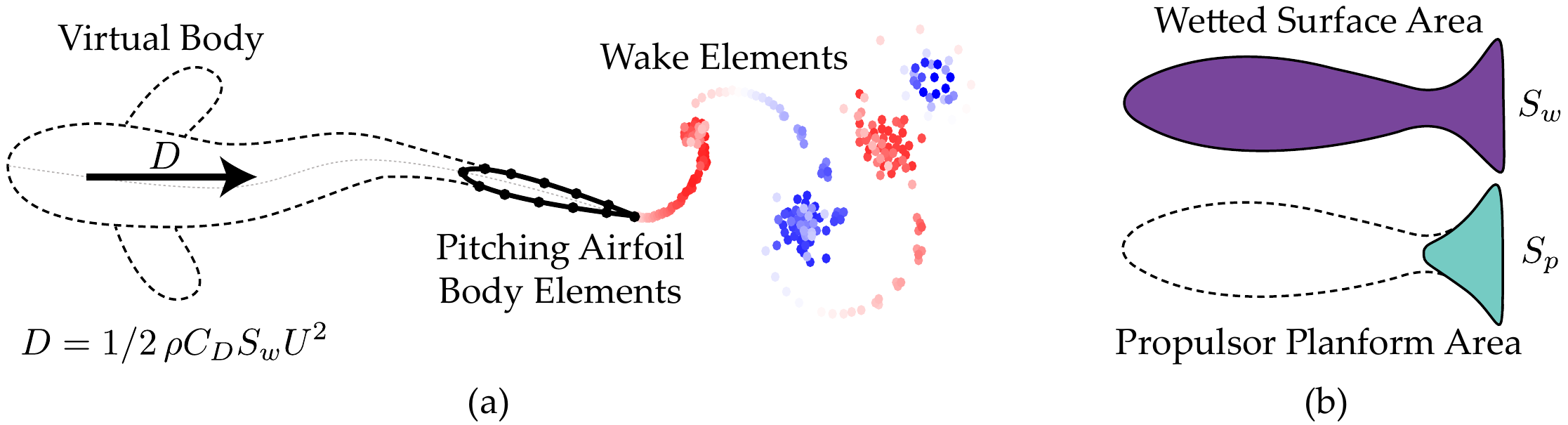}
		}	
	\end{center}
	\caption{(a) Illustration of the potential flow method.  The presence of a virtual body acts as a drag force on a two-dimensional thrust-generating pitching airfoil.  On the airfoil the circles designate the endpoints of the doublet and source body elements.  The circles in the wake designate the end points of the doublet wake elements.  Positive and negative vorticity in the wake are shown as red and blue, respectively. (b) Side view of a generic fish showing the wetted surface area and propulsor planform for area.}
	\label{fig:meth_sch} 
\end{figure}

Computations are performed on an idealized swimmer that is a combination of a virtual body and a two-dimensional airfoil pitching about its leading edge (Figure \ref{fig:meth_sch}a).  The virtual body is not present in the computational domain except as a drag force, $D$, acting on the airfoil.  A drag law is specified that depends on the swimming speed in the following manner,
\begin{align}\label{eq:draglaw}
D = 1/2 \, \rho C_D S_w U^2,
\end{align}

\noindent where $\rho$ is the fluid density, $C_D$ is a drag coefficient, $S_w$ represents the wetted surface area of a swimmer (Figure \ref{fig:meth_sch}b) including its body and propulsor and $U$ is the swimming speed.  This drag force represents the total drag acting on the body and fin of a streamlined swimmer and the pitching airfoil represents its thrust-producing fin.  A drag law that follows a $U^2$ scaling is a reasonable approximation of the drag scaling for steady streamlined bodies operating at high Reynolds numbers, that is $Re \geq \mathcal{O}(10^6)$ \cite[]{Munson1998}.  A $U^{3/2}$ Blasius scaling for laminar boundary layers could also be used \cite[]{Landau1987a} to model swimming at lower Reynolds numbers.  In fact, the transition between a high Reynolds number drag law and a Blasius drag law has been noted to occur at lower $Re$ in fish swimming than in steady flow cases around $Re = \mathcal{O}(10^4)$ determined from a wide range of biological data \cite[]{Gazzola2014}.  The high Reynolds number drag law (Eq. \ref{eq:draglaw}) will be used to model the presence of a drag-producing body in the numerical simulations.  Later, the scalings that are developed for high Reynolds numbers will be generalized to account for a Blasius drag law as well.

The virtual body and `fin' have several properties and parameters including the prescribed drag law, drag coefficient, mass, wetted area, fin chord and fin shape.    Four drag coefficients are chosen (see Table \ref{tab:parameters}) in order to cover a range typical of biology \cite[]{Lighthill1971,Fish1998a}.  High drag coefficients represent a poorly streamlined body and \textit{vice versa}.  Three non-dimensional body masses are also specified by normalizing the body mass by the added mass of the airfoil propulsor,
\begin{align}
m^* \equiv \frac{m}{\rho S_p c}.
\end{align}

\noindent The pitching airfoil has a chord length of $c = 1\, \mbox{m}$ and a NACA 0012 cross-sectional shape \cite[]{Jones1997,Mackowski2015}.  Even though this is a two-dimensional study the propulsor planform area is specified as $S_p$, which is the chord length multiplied by a unit span length.  The non-dimensional mass, $m^*$, affects not only the acceleration of a swimmer due to net time-averaged forces over a period, but also the magnitude of the surging oscillations that occur within a period due to the unsteady forcing of the pitching airfoil.  We further non-dimensionalize the wetted area,
\begin{align}
S_{wp} \equiv \frac{S_w}{S_p}\dbq{,}
\end{align}

\noindent and specify a range of five area ratios for the simulations.  When the area ratio is large this represents a large body connected to a small caudal fin.  The minimum value for the area ratio is $S_{wp} = 2$, which is the case when there is no body and only an infinitely-thin propulsor.  Throughout this study, it will become clear that an important nondimensional combination of parameters is,
\begin{align}
Li \equiv C_D\, S_{wp}\dbq{.}
\end{align}

\noindent Here we define the Lighthill number, $Li$, which was first defined in \cite{Eloy2012a} in a slightly different way.  The Lighthill number characterizes how the body and propulsor geometry affects the balance of thrust and drag forces on a swimmer.  For example, for a fixed thrust force and propulsor area, the self-propelled swimming speed scales as $U \propto Li^{-1/2}$.  If $Li$ is high then a swimmer produces high drag at low speeds, leading to low self-propelled swimming speeds and \textit{vice versa}.  In the present study, the Lighthill number covers a range that is typical of biology, $0.01 \leq Li \leq 1$ \cite[]{Eloy2012a}.

Finally, the kinematic motion is parameterized with a pitching frequency, $f$, and a peak-to-peak trailing-edge amplitude, $A$, reported as a non-dimensional amplitude-to-chord ratio, \begin{align}
\As = A/c.
\end{align}  

\noindent The amplitude-to-chord ratio is related to the maximum pitch angle, that is, $\theta_0 = \mbox{sin}^{-1}\left(\As/2\right)$.   The airfoil pitches about its leading edge with a sinusoidal motion described by $\theta(t) = \theta_0 \, \mbox{sin} \left( 2 \pi f t\right)$.  
\begin{table}
   \begin{center}
	\begin{tabular}{cccccccc} 
         $C_D$ & & 0.005 & 0.01 & 0.05 & 0.1 & |&    \\
         $m^*$& & 2 & 5 & 8 & |  & | &    \\
         $S_{wp}$& & 2 & 4 & 6 & 8 & 10 &   \\
	$f$  (Hz)& & 0.1 & 0.5 & 1 & 5 & 10 &   \\
	$\As$& & 0.25 & 0.375 & 0.5 & 0.625  & 0.75 &    \\
	$\theta_0$  (deg.)& & 7.2  & 10.8  & 14.5 &  18.2 &  22 & \\
	\end{tabular} 
   \end{center}
  	\caption{Simulations parameters used in the present study.}
 	\label{tab:parameters}
\end{table}

All of the input parameters used in the present simulations are reported in Table \ref{tab:parameters}.  The  combinatorial growth of the simulation parameters produces 1,500 simulations.  All of these simulations data are presented in Figures \ref{perform}--\ref{selfpropscale}.

\subsection{Numerical Methods}
To model the flow over the foil, an unsteady two-dimensional potential flow method is employed where the flow is assumed to be irrotational, incompressible and inviscid.  We follow \cite{Katz2001} and \cite{Quinn2014}, in that the general solution to the potential flow problem is reduced to finding a distribution of doublets and sources on the foil surface and in the wake that satisfy the no flux boundary condition on the body at each time step.  The elementary solutions of the doublet and source both implicitly satisfy the far-field boundary condition.  We use the Dirichlet formulation to satisfy the no-flux condition on the foil body.  To solve this problem numerically, the singularity distributions are discretized into constant strength line boundary elements over the body and wake.   Each boundary element is assigned one collocation point within the body where a constant potential condition is applied to enforce no flux through the element.  This results in a matrix representation of the boundary condition that can be solved for the body doublet strengths once a wake shedding model is applied.  At each time step a wake boundary element is shed with a strength that is set by applying an explicit Kutta condition,  where the vorticity at the trailing edge is set to zero so that flow leaves the airfoil smoothly \cite[]{Willis2006,Zhu2007a,Wie2009,Pan2012}.

At every time step the wake elements advect with the local velocity such that the wake does not support any forces \cite[]{Katz2001}.  During this wake rollup, the ends of the wake doublet elements, which are point vortices, must be de-singularized for the numerical stability of the solution \cite[]{Krasny1986}.  At a cutoff radius of $\epsilon/c = 5 \times 10^{-2}$, the irrotational induced velocities from the point vortices are replaced with a rotational Rankine core model.  The tangential perturbation velocity over the body is found by a local differentiation of the perturbation potential.  The unsteady Bernoulli equation is then used to calculate the pressure acting on the body.  The airfoil pitches sinusoidally about its leading edge, and the initial condition is for the airfoil trailing edge to move upward at $\theta = 0$.

In this study, only the streamwise translation of the self-propelled swimmer is unconstrained while the other degrees-of-freedom follow fully prescribed motions. To calculate the position and speed of the swimmer the equations of motion are then solved through a one-way coupling from the fluid solution to the body dynamics.   Following \cite{Borazjani2008a}, the loose-coupling scheme uses the body position and velocity at the current $n^{th}$ time step to explicitly solve for the position and velocity at the subsequent $(n+1)^{th}$ time step,
\begin{align} \label{EoM}
x_{LE}^{n+1} &= x_{LE}^{n} + \frac{1}{2}\left( U^{n+1} + U^{n} \right) \Delta t \\
U^{n+1} &= U^{n} + \frac{F_{x,net}^n}{m} \Delta t
\end{align}

\noindent Here the time step is $\Delta t$, the net force acting on the swimmer in the streamwise direction is $F_{x,net}$, and the $x$-position of the leading edge of the airfoil is $x_{LE}$.

\subsection{Output Parameters}
There are several output parameters used throughout this study.  For many of them, we examine their mean values time-averaged over an oscillation cycle, which are denoted with an overbar such as $\overline{(\boldsymbol{\cdot})}$.  Mean quantities are only taken after a swimmer has reached the steady-state of its cycle-averaged swimming speed.  For instance, when this occurs the steady-state cycle-averaged swimming speed will be described as the mean swimming speed and denoted as $\overline{U}$.  Additionally, the mean swimming speed will also be reported as a nondimensional stride length,
\begin{align} \label{nonDspeed}
U^* \equiv \frac{\overline{U}}{fc}
\end{align}

\noindent This represents the distance travelled by a swimmer in chord lengths over one oscillation cycle and it is the inverse of the reduced frequency.  The reduced frequency and the Strouhal number, 
\begin{align} \label{nonDfreq}
k \equiv \frac{f c}{ \overline{U}} && St \equiv \frac{f A}{ \overline{U}}
\end{align}

\noindent are two parameters that are typically input parameters in fixed-velocity studies but become output parameters in self-propelled studies since the mean swimming speed is not known \textit{a priori}.  The reduced frequency represents the time it takes a fluid particle to traverse the chord of an airfoil compared to the period of oscillation.  For high reduced frequencies the flow is highly unsteady and it is dominated by added mass forces and for low reduced frequencies the flow can be considered quasi-steady where it is dominated by circulatory forces.  The Strouhal number can be interpreted as the cross-stream distance between vortices in the wake compared to their streamwise spacing.  For a fixed amplitude of motion, when the Strouhal number is increased the vortices in the wake pack closer to the trailing edge and have a larger influence on the flow around an airfoil.  The time-averaged thrust and power coefficients depend upon the reduced frequency and the Strouhal number and are, 
\begin{align} \label{coeffs}
C_T \equiv \frac{\overline{T}}{ \rho S_p f^2 A^2} && C_P \equiv \frac{\overline{P}}{ \rho S_p f^2 A^2 \overline{U}}
\end{align}

\noindent These coefficients are nondimensionalized with the added mass forces and added mass power from small amplitude theory \cite[]{Garrick1936}.  Also, the mean thrust force is calculated as the time-average of the streamwise directed pressure forces and the time-averaged power input to the fluid is calculated as the time average of the negative inner product of the force vector and velocity vector of each boundary element, that is, $P = -\int_\mathcal{S} \mb{F}_{\text{ele}} \cdot \mb{u}_{\text{ele}} \, d\mathcal{S}$ where $\mathcal{S}$ is the body surface.  The ratio of these coefficients leads to the propulsive efficiency, $\eta$, which is intimately linked to the swimming economy, $\xi$, and the cost of transport, $CoT$,
\begin{align} \label{ener}
\eta \equiv \frac{\overline{T}\overline{U}}{\overline{P}}=\frac{C_T}{C_P} && \xi \equiv \frac{\overline{U}}{\overline{P}} && CoT \equiv \frac{\overline{P}}{m \overline{U}}
\end{align}

\noindent The propulsive efficiency is the ratio of useful power output to the power input to the fluid.  In self-propelled swimming we define this quantity in the potential flow sense \cite[]{Lighthill1971}, that is, the mean thrust force is calculated as the integration of the pressure forces alone.  In this sense the propulsive efficiency is not ill-defined for self-propelled swimming, however, this definition is typically unused in experimental studies since, until recently, it has been difficult to resolve the pressure forces alone acting on a body in an experiment \cite[]{Lucas2016}.  The swimming economy is a measure of the distance travelled per unit energy; it is effectively a `miles-per-gallon' metric.  While the transfer of power input into swimming speed is intimately linked to propulsive efficiency, the economy reflects the fact that even if the efficiency is constant it still takes more power to swim faster, such that, in general, $\xi$ decreases as the swimming speed increases.  The cost of transport first made its appearance in an article from \cite{Gabrielli1950} as the tractive force to weight ratio and later it was re-introduced as the cost of transport \cite[]{Schmidt-Nielsen1972,Tucker1975}.  The cost of transport is the inverse of the swimming economy that is reported on a per unit mass basis.  It is widely used throughout biological literature \cite[]{Videler1993} and is a useful engineering metric since its inverse is the proportionality constant between the range of a vehicle, $\mathcal{R}$, and its energy density ($\mathcal{E} \equiv \mbox{energy per unit mass} = E/m$), that is,
\begin{align} \label{range}
\mathcal{R} = \left(\frac{1}{CoT}\right) \mathcal{E}
\end{align}

\noindent For example, the energy density of various battery technologies is relatively constant as the battery size is scaled \cite[]{Ding2015}.  Then in this case, the $CoT$ directly connects the range of a vehicle with the current state-of-the-art in battery technology.  The $CoT$ and $\xi$ are also readily measurable in self-propelled experiments \cite[]{Moored2011b} without measuring the three-dimensional time-varying flow \cite[]{Dabiri2005}.  In addition, both energy metrics can be connected to the efficiency such that $CoT = \overline{D}/m\eta$ and $\xi =\eta/ \overline{D}$, by considering that the time-averaged thrust and drag must balance in non-accelerating self-propelled swimming.  The time-averaged drag in this study is directly given in Eq. (\ref{eq:draglaw}).  Even in experiments where the drag law is not specified a nondimensional cost-of-transport,
\begin{align} \label{nonDCoT}
CoT^* = \frac{CoT\, m}{1/2\, \rho S_w \overline{U}^2},
\end{align}

\noindent can be used as discussed in \cite{Fish2016}.  The $CoT^*$ reflects an inverse relationship with the propulsive efficiency if the normalizing characteristic drag force is properly chosen.
\begin{figure}
\vspace{0mm}
	\begin{center}
	\centerline{ 		{
		\includegraphics[width=0.99\textwidth]{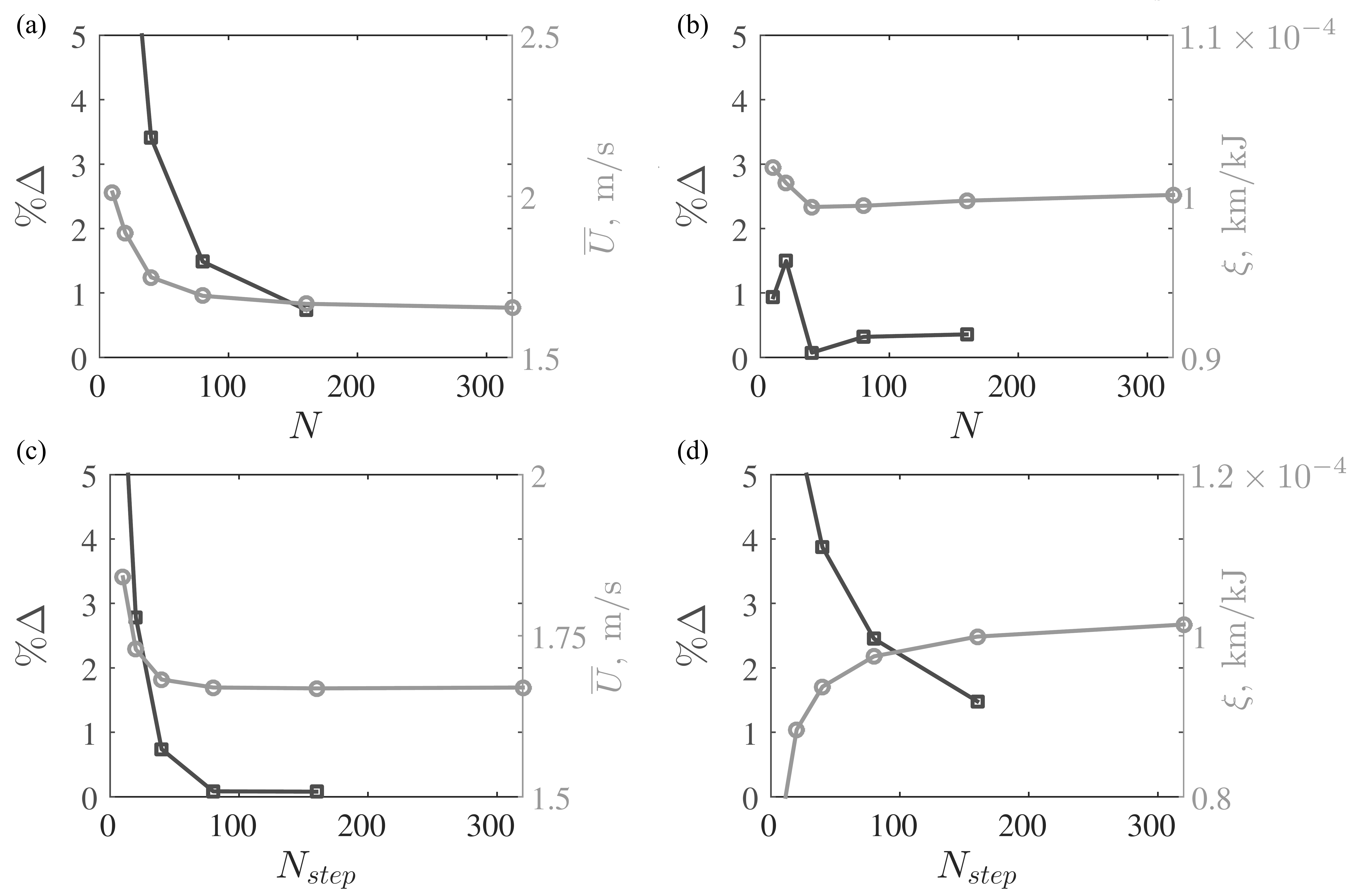}}
			}
	\end{center}
	\caption{Solution dependence as a function of the number of body elements (top row) and the number of time steps per cycle (bottom row).  The left axes represent the percent change in the solution when the number of elements is doubled.  The right axes represent the value of the solution.  The left subfigures present the mean swimming speed.  The right subfigures present the swimming economy.}
	\label{converge}
\end{figure}

Lastly, throughout this study the various output quantities calculated from the numerical simulations will be compared with their predicted values from scaling relations.  This comparison will form normalized quantities denoted by an over hat, such as,
\begin{align} \label{nonDspeed}
\hat{(\boldsymbol{\cdot})} \equiv \frac{(\boldsymbol{\cdot})_{Simulation}}{(\boldsymbol{\cdot})_{Predicted}}
\end{align}

\noindent If these normalized quantities are equal to one then the related scaling relation perfectly predicts their value.

\subsection{Discretization Independence}
Convergence studies found that the mean swimming speed and swimming economy changed by less than 2\% when the number of body elements $N$ ($= 150$) and the number of times steps per cycle $N_t$ ($ = 150$) were doubled (Figure \ref{converge}).  The parameters for the convergence simulations are $C_D = 0.05$, $m^* = 5$, $S_{wp} = 6$, $f = 1$ Hz and $A/c = 0.5$.  The simulations were started with an initial velocity prescribed by the scaling relations summarized in Table \ref{tab:scalings} and run for fifty cycles of swimming.  The time-averaged data are obtained by averaging over the last cycle.  The numerical solution was validated using canonical steady and unsteady analytical results (see \cite{Quinn2014}).  For all of the data presented, $N = 150$ and $N_t = 150$.

\subsection{Linear Unsteady Airfoil Theory} \label{sec:Garrick}
\cite{Garrick1936} was the first to develop an analytical solution for the thrust production, power consumption and efficiency of a sinusoidally heaving and pitching airfoil or hydrofoil.  This theory extended Theodorsen's theory \cite[]{Theodorsen1935} by accounting for the singularity in the vorticity distribution at the leading-edge in order to calculate the thrust and by determining the power consumption.  Both approaches are \textit{linear} theories and assume that the flow is a potential flow, the hydrofoil is infinitesimally thin, that there are only small amplitudes of motion and that the wake is planar and non-deforming.  In the current study Garrick's solution for the thrust coefficient and power coefficient will be assessed for it's capability in predicting the data generated from the \textit{nonlinear} boundary element method simulations where the only assumption is that the flow is a potential flow.  In this way, the potential flow numerical solutions are able to directly probe how the nonlinearities introduced by \textit{large-amplitude motions} and by \textit{nonplanar and deforming wakes} affect the scaling relations predicted by linear theory.  Garrick's (\citeyear{Garrick1936}) exact solutions for the thrust and power coefficients of a hydrofoil pitching about its leading edge are
\begin{align} 
C^G_{T}(k) &= \frac{3\pi^3}{32} \, - \frac{\pi^3}{8} \, \left[ \frac{3F}{2} - \frac{G}{2\pi k} + \frac{F}{\pi^2 k^2} - \left( F^2 + G^2 \right)\left( \frac{1}{\pi^2 k^2} + \frac{9}{4} \right) \right] \label{eq:thrust}\\
C^G_{P}(k) &= \frac{3\pi^3}{32} + \frac{\pi^3}{16}\left[ \frac{3F}{2} + \frac{G}{2 \pi k}\right]. \label{eq:power}
\end{align}

\noindent Here $F$ and $G$ are the real and imaginary parts of the Theodorsen lift deficiency function, respectively \cite[]{Theodorsen1935}, and as shown in (\ref{coeffs}) the thrust and power coefficients are normalized by the added mass forces and added mass power, respectively.  Note that the thrust equation for a pitching airfoil presented in \cite{Garrick1936} has an algebraic error as first observed by \cite{Jones1997}.

Garrick's theory can be decomposed into its added mass and circulatory contributions, which in equations (\ref{eq:thrust}) and (\ref{eq:power}) are the first terms and the second terms (denoted by the square brackets), respectively.  The added mass thrust contribution, the circulatory thrust contribution, and the total thrust are graphed in Figure \ref{thrust}a.  It can be observed that the added mass thrust contribution is always positive (thrust producing) and the circulatory thrust contribution is always negative (drag inducing) for pitching about the leading edge.  The total power solution is graphed in Figure \ref{power}a.  These linear unsteady airfoil theory solutions will be assessed for their capability in predicting the scaling of the thrust and power of the nonlinear numerical data presented in this study.

\section{Results} \label{sc:results}
\begin{figure}
\vspace{0mm}
	\begin{center}
	\centerline{ 		{
		\includegraphics[width=0.99\textwidth]{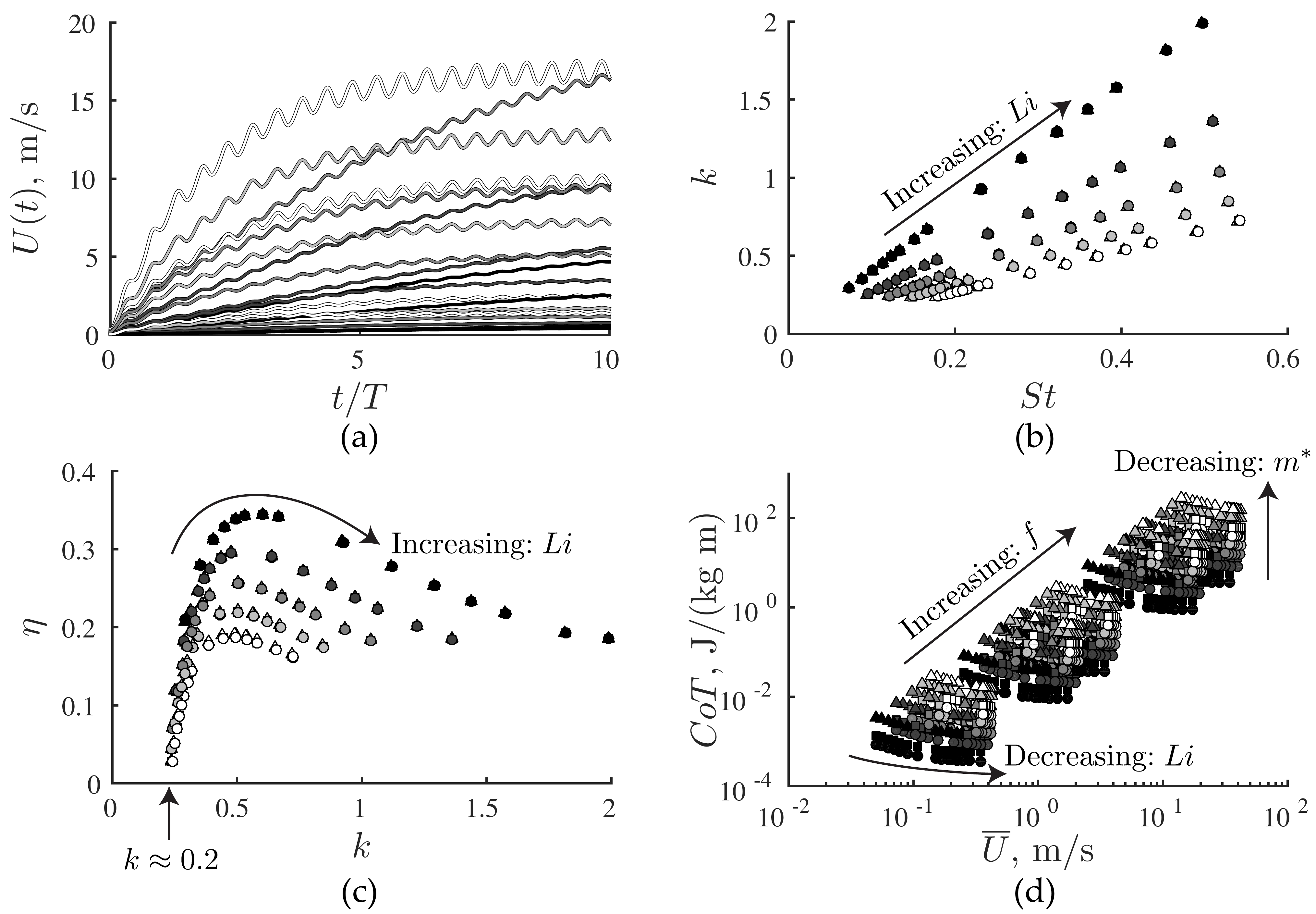}}
			}
	\end{center}
	\caption{(a) Time-varying swimming speed.  (b) Range of reduced frequencies and Strouhal numbers for the simulations.  (c) Propulsive efficiency as a function of reduced frequency.  (d) Cost-of-transport as a function of the mean swimming speed.  The line and marker colors indicate the amplitude of motion of a simulation where going from the smallest to the largest amplitude is mapped from black to white, respectively.  The marker style also denotes the nondimensional mass with the triangle, square and circle indicating $m^* = 2, \, 5,$ and $8$, respectively.}
	\label{perform}
\end{figure}

Figure \ref{perform} presents the data from 1,500 simulations produced through the combinatorial growth of the simulation parameters in Table \ref{tab:parameters}.  The line and marker colors (Figure \ref{perform}) indicate the amplitude of motion where going from the smallest to the largest amplitude is mapped from black to white, respectively.  The marker style also denotes the nondimensional mass with a triangle, square and circle indicating $m^* = 2, \, 5,$ and $8$, respectively.

Figure \ref{perform}a shows the time-varying swimming speed of a subset of swimmers that were given an initial velocity that was $\mathcal{O}(2\%)$ of their mean speed.  The swimmers start from effectively rest and accelerate up to their mean swimming speed where their instantaneous speed fluctuates about the mean due to the unsteady forces acting on the pitching airfoil.  As the amplitude of motion is increased the number of cycles to reach a mean speed is decreased, and the amplitude of the velocity fluctuations is increased.  Although not obvious from the figure, as $m^*$ is increased the velocity fluctuations of the swimmer are reduced and the number of cycles to reach its mean speed is increased.

For all subsequent data presented in this study these preliminary simulations (Figure \ref{perform}a) were used with the scaling laws developed later to estimate the steady-state value of their cycle-averaged swimming speed.  These estimates were then used to seed the initial velocity of the swimmers to within $\mathcal{O}(10\%)$ of their actual mean swimming speeds.  Then this second set of simulations were run for 75 cycles of swimming and the net thrust coefficient, $C_{T,net} = \overline{T}_{net}/(1/2 \, \rho S_w \overline{U}^2)$\dbq{,} at the end of the simulations was less than $C_{T,net} < 2\times10^{-3}$, indicating that the mean swimming speed had been effectively reached.  Here, the net thrust is the difference between the thrust from pressure forces and the drag from the virtual body, that is $\overline{T}_{net} = \overline{T} - \overline{D}$.

Figure \ref{perform}b shows that the combination of input parameters leads to a reduced frequency range of $0.24 \leq k \leq 2.02$ and a Strouhal number range of $0.073 \leq St \leq 0.54$.  These ranges cover the parameter space typical of animal locomotion \cite[]{Taylor_2003} and laboratory studies \cite[]{Quinn2015} alike.  The reduced frequency and Strouhal number are directly linked by the definition
\begin{align} \label{eq:k_vs_St}
k = \frac{St}{\As}.
\end{align}

\noindent The linear relationship between $k$ and $St$ and the inverse relationship between $k$ and $\As$ can be observed in the figure, which are merely a consequence of their definitions.  As $Li$ is increased the drag forces acting on the swimmer are increased leading to a lower mean swimming speed when all other parameters are held fixed.  This, in turn, increases $k$ and $St$ of the swimmer.  Variation in the frequency of motion and $m^*$ have no effect and negligible effect, respectively, on $k$ and $St$ of a swimmer, at least for the range of $m^*$ investigated.  The scaling relations proposed later will capture these observations.  

The propulsive efficiency as a function of the reduced frequency is shown in Figure \ref{perform}c.  For a fixed amplitude of motion and an increasing $Li$, the efficiency climbs from a near zero value around $k \approx 0.2$ to a peak value around $0.5 < k < 0.6$ after which $\eta$ decreases with increasing $k$.  It should be noted that the end of each efficiency curve of fixed $\As$ is at the same $Li$.  This efficiency trend has been observed in numerous studies with the transition from thrust to drag and the subsequent rise in efficiency being attributed to a competition between thrust production and viscous drag acting on the airfoil \cite[]{Quinn2015,Das2016}.  Here, however, the rise in efficiency found in an inviscid domain is due to a competition between the induced drag from the presence of the wake and the thrust produced by the pitching motions.  This point will be discussed further in the section on the thrust and power scaling relations.  As the pitching amplitude is increased there is a monotonic decrease in the efficiency.  This trend is observed in experiments \cite[]{Buchholz2008} and viscous flow simulations \cite[]{Das2016} for angular amplitudes greater than $\theta_0 \approx 8^o$, but the trend reverses for smaller amplitudes.  The discrepancy is likely due to the definition of $\eta$ used in those studies as compared to this study.  In those cases, the efficiency is based on the net thrust force, which for small amplitudes is dominated by the viscous drag on the airfoil.  This then strongly attenuates the efficiency calculation at small amplitudes, but has less of an effect as the amplitude is increased.  The peak efficiencies calculated in this study range from 19\%--34\% for the highest and lowest amplitudes, respectively.  In previous pitching studies, the peak efficiency was measured at 21\% \cite[]{Buchholz2008} for a flat plate and calculated at 16\% \cite[]{Das2016} for a NACA 0012 airfoil, which is indeed similar in magnitude to the high-amplitude cases presented here.  The efficiency will also be slightly overpredicted by using an explicit Kutta condition, as is the case in our numerical implementation, as compared to an implicit Kutta condition \cite[]{Basu1978,Jones1997}.  

Varying the frequency has no effect on the reduced frequency and Strouhal number, which in turn leads to no effect on the efficiency.  This trend has been elucidated across a wide variation of aquatic animals at least when $Re \geq \mathcal{O}(10^4)$ \cite[]{Gazzola2014}.  Additionally, these data indicate that in self-propelled swimming for a fixed amplitude of motion, controlling $Li$ is crucial in controlling $k$ and $St$ of a swimmer and, subsequently, in maximizing the efficiency. 

The parameter space of the current study covers over three orders of magnitude in mean swimming speed and over six orders of magnitude in cost of transport as shown in Figure \ref{perform}d.  As the frequency is increased both the mean speed and $CoT$ are increased as expected.  As $Li$ is decreased, the lower relative drag results in a higher mean speed and a moderate reduction in $CoT$.  Finally, as $m^*$ is decreased the $CoT$ is increased with a small effect on the mean speed.  This is purely an implication of the definition of $CoT$, which is inversely proportional to the mass.  In this study $Li$ and $m^*$ have been varied independently, which is possible in engineered vehicles, however, in animals it would be reasonable to assume that $m^*$ is proportional to $S_{wp}$ and in turn $Li$.  Then, for this case, there would be a more dramatic decrease in $CoT$ with a decrease in $Li$.
\kwm{\begin{figure}
\vspace{0mm}
	\begin{center}
	\centerline{ 		{
		\includegraphics[width=0.99\textwidth]{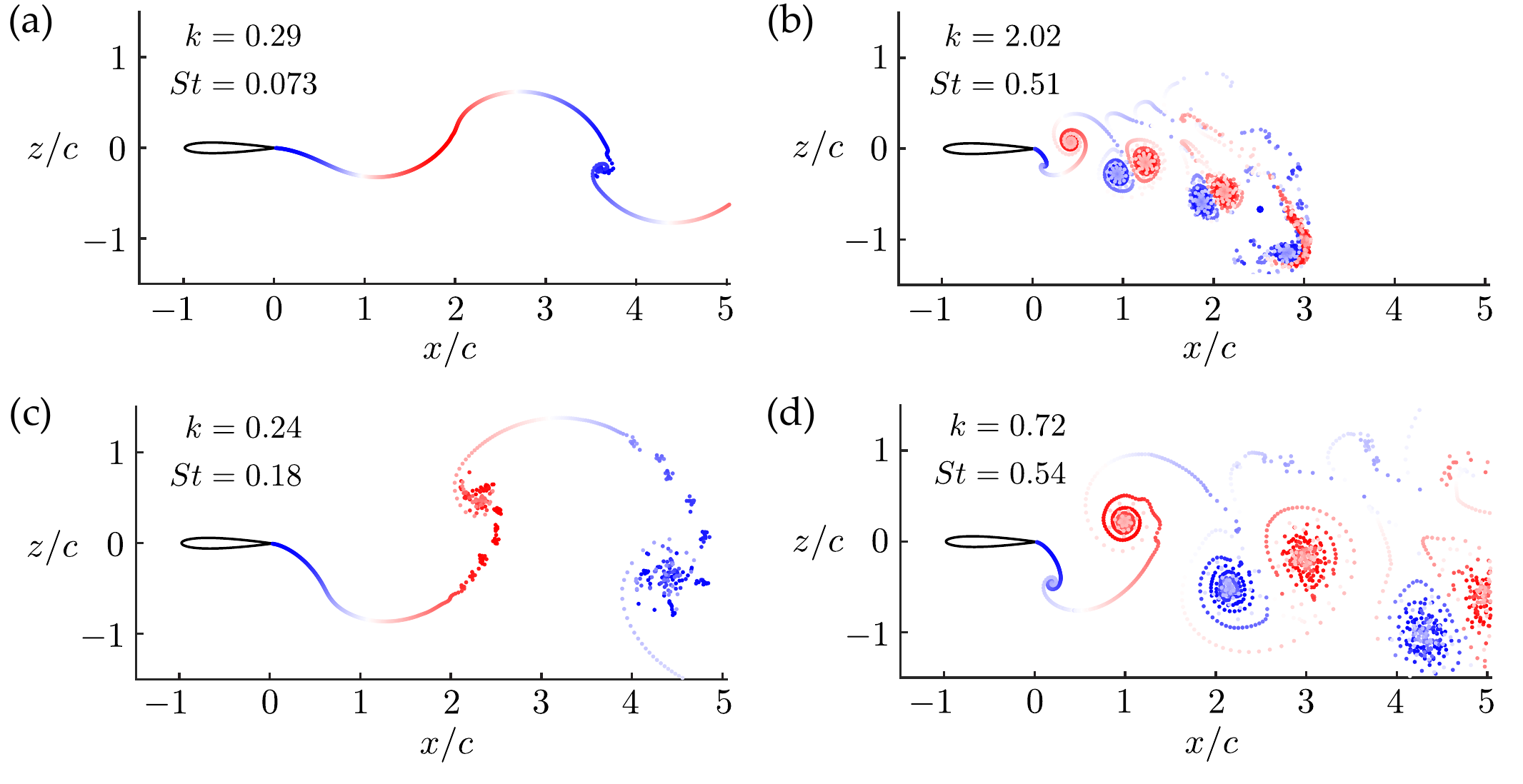}}
			}
	\end{center}
	\caption{Vortical wake structures for (a) $Li = 0.01$, $A^* = 0.25$, (b) $Li = 1$, $A^* = 0.25$,  (c) $Li = 0.01$, $A^* = 0.75$,  and (d) $Li = 1$, $A^* = 0.75$.  The circles in the wakes designate the end points of the doublet wake elements.  Positive and negative vorticity in the wakes are shown as red and blue, respectively.}
	\label{wake}
\end{figure}}

As will be discussed in Section \ref{sec:summary} the \textit{nondimensional performance} of a self-propelled two-dimensional pitching airfoil is completely defined by two nondimensional variables, namely, the Lighthill number, $Li$, and the amplitude-to-chord ratio, $\As$.  In order to provide physical insight for a scaling analysis, the vortical wake structures for the four combinations of the lowest and highest $Li$ and $A^*$ are presented in Figure \ref{wake}.  Each combination produces a different $k$ and $St$, which are labeled on the subfigures. All of the cases show the formation of a reverse \vK vortex street with two vortices shed per oscillation cycle.  In the lowest $St$ case (Figure \ref{wake}a, $Li = 0.01$ and $A^* = 0.25$) the wake deforms away from the $z/c = 0$ line even for this low amplitude of motion.  However, the roll up of the vortex wake is weak and is not evident until $x/c = 3$.  Figure \ref{wake}b ($Li = 1$ and $A^* = 0.25$) presents the highest $k$ case with a high $St$, which leads to a markedly stronger roll up of the wake.  In fact, the trailing edge vortex is observed to be forming nearly overtop of the trailing edge.  The high $St$ also leads to a compression of the wake vortices in the streamwise direction, which enhances the mutual induction between vortex pairs \cite[]{Marais2012a}.  Consequently, the wake deflects asymmetrically as observed in both experiments \cite[]{Godoy-diana2008} and direct numerical simulations \cite[]{Das2016}.  Figure \ref{wake}c ($Li = 0.01$ and $A^* = 0.75$) is the lowest $k$ case, but with a moderate $St$ where coherent vortices are formed closer to the airfoil than in Figure \ref{wake}a around $x/c = 2$.  The shear layers feeding the vortices are observed to be broken up by a Kelvin-Helmholtz instability before entraining into the vortex cores as observed in experiments \cite[]{Quinn2014}. Figure \ref{wake}d ($Li = 1$ and $A^* = 0.75$) shows the highest $St$ case with a moderate $k$ where again a deflected wake is observed.  In comparison the to the low $A^*$, high $Li$ case (Figure \ref{wake}b) the trailing-edge vortex is again observed to be forming expediently, yet the vortex is forming further away from the airfoil in the cross-stream direction than Figure \ref{wake}b.  This observation has implications on the power scaling relation discussed in the subsequent section.

\section{Thrust and Power Scaling Relations}\label{sec:scaling}
\begin{figure}
\vspace{0mm}
	\begin{center}
	\centerline{ 		{
		\includegraphics[width=0.99\textwidth]{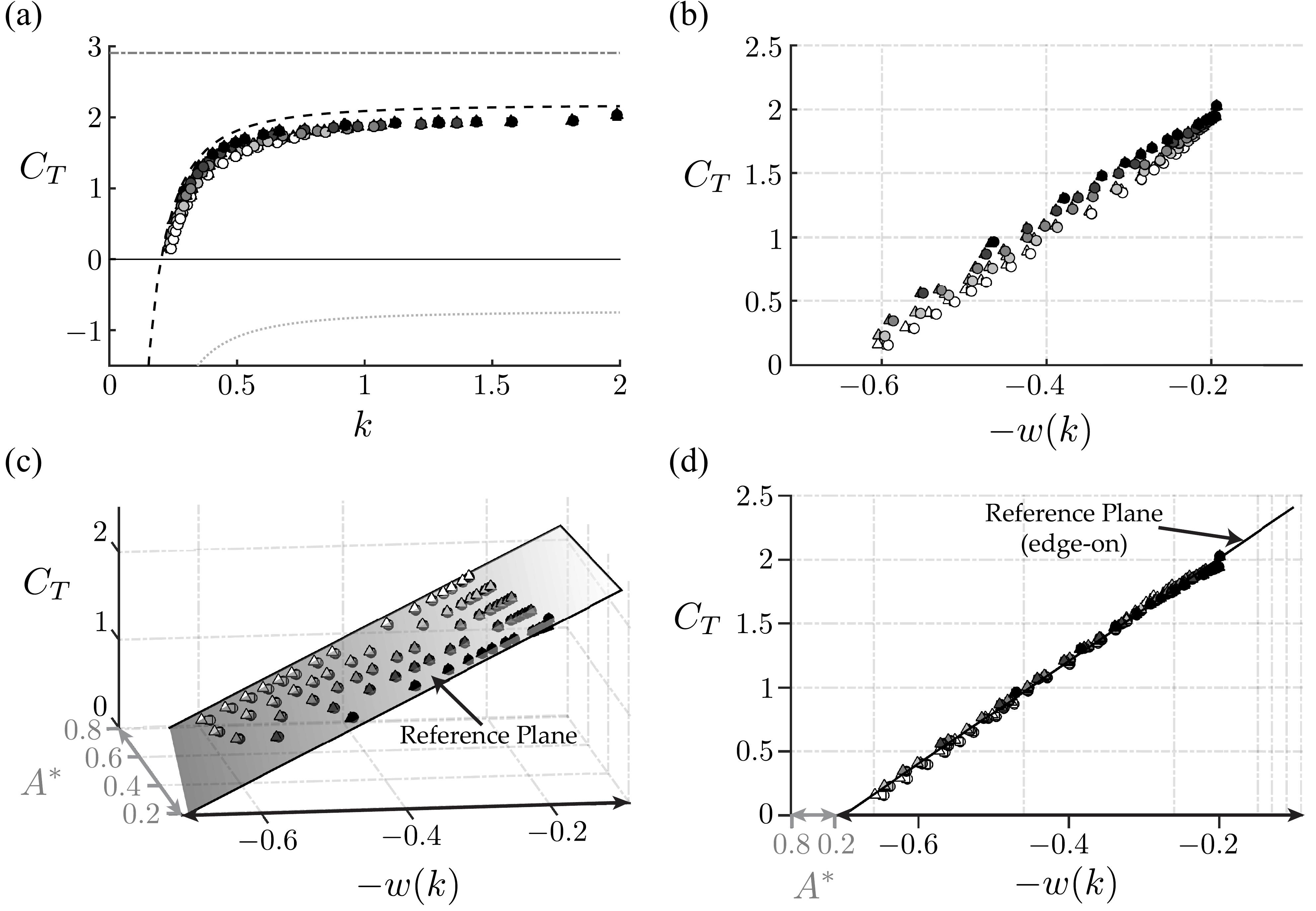}}
			}
	\end{center}
	\caption{(a) Log-log graph of the time-averaged thrust as a function of frequency. (b) Markers indicate the thrust coefficient as a function of reduced frequency from the simulations.  The marker colors indicate the amplitude of motion, which goes from $\As = 0.25$--$0.75$ with a gradient of color from black to white, respectively.  The black dashed line represents Garrick's theory.  The dark grey dash-dot and light grey dotted lines represent added mass thrust-producing forces and circulatory drag-inducing forces, respectively, from Garrick's theory.}
	\label{thrust}
\end{figure}

Figure \ref{thrust}a presents the time-averaged thrust coefficient as a function of the reduced frequency, and it shows that the data nearly collapse to a curve that asymptotes to a constant at high reduced frequencies.  This collapse occurs when the thrust force only includes pressure forces and does not include viscous skin friction drag.  Previously, \cite{Dewey_2013} and \cite{Quinn2014} considered that the thrust forces from a pitching panel and airfoil scale with the added mass forces, that is, $\overline{T} \propto \rho S_p f^2 A^2$, which does indeed lead to well-collapsed data.  However, this scaling only captures the asymptotic value and does not account for the observed variation in the thrust coefficient with the reduced frequency, nor does the scaling indicate that the data should collapse well when plotted as a function of reduced frequency.  Additionally, the previous scaling relation can only lead to a reasonable prediction of thrust and, consequently, the self-propelled swimming speed when the reduced frequency is high.  To overcome these limitations both the added mass and circulatory forces must be considered. 

For low reduced frequencies, circulatory forces dominate over added mass forces.   In this quasi-steady regime, vorticity is continually shed into the wake with a time-varying sign of rotation.  The proximity of the wake vorticity produces a time-varying upwash and downwash, which modifies the bound circulation and induces streamwise forces that are proportional to the induced angle of attack.  Garrick's theory (dashed black line in Figure \ref{thrust}a) exactly accounts for the additional circulation and the induce drag of the wake vorticity \cite[]{Garrick1936}.  Indeed, when Garrick's theory is plotted, the numerical data is seen to follow the same trend with the theory slightly over-predicting the thrust coefficient data.  Similarly, this discrepancy has been observed in other studies \cite[]{Jones1997,Mackowski2015}. 

Garrick's theory can be decomposed into its added mass and circulatory contributions, which are graphed in Figure \ref{thrust}a.  The decomposition shows that the total thrust is a trade-off between the added mass forces that are thrust-producing and the circulatory forces that are drag-inducing.  Both the added mass and circulatory terms are combined to form the exact Garrick relation,
\begin{align} \label{CT_Garrick}
C^G_{T}(k) &= c_1 \, - c_2 \, w(k) \\
 \mbox{where } w(k) &= \frac{3F}{2} + \frac{F}{\pi^2 k^2} - \frac{G}{2\pi k} - \left( F^2 + G^2 \right)\left( \frac{1}{\pi^2 k^2} + \frac{9}{4} \right). \nonumber 
\end{align} 

\noindent Here the wake function $w(k)$ is a collection of the reduced frequency dependent terms that in equation (\ref{eq:thrust}) are within the square brackets.  The coefficients have exact values from theory of $c_1 = 3 \pi^3/32$ and $c_2 = \pi^3/8$. However, to more accurately model pitching airfoils that do not adhere to the theory's assumptions (detailed in Section \ref{sec:Garrick}) these coefficients are left to be determined.

If the Garrick thrust relation accurately predicts the scaling behavior of the nonlinear numerical data then (\ref{CT_Garrick}) suggests that the numerical thrust coefficient data should collapse to a line if it is graphed against $w(k)$.  Indeed, good collapse of the data to a line is observed in Figure \ref{thrust}b, however, the thrust coefficient is observed to show a mild dependency on the amplitude of motion.   It is observed that an increase in $\As$ results in a decrease in the thrust coefficient.  Even though Garrick's theory accounts for the induced drag from the wake vortex system, it does not account for the form drag due to the shed vortices at the trailing-edge, which is implicitly incorporated in the numerical solution.  Consider that the induced drag term in (\ref{CT_Garrick}) is only a function of the reduced frequency, not the amplitude of motion, and as such it is always present even for an infinitesimally thin body with infinitesimally small motions.  On the other hand, it is well-known that for bodies of finite thickness or inclined at a finite angle, vortex shedding at the trailing-edge can lead to form drag.


To correct the small amplitude theory, we can consider for a fixed reduced frequency and a fixed chord length that the streamwise spacing of wake vortices (proportional to $U/f$) will be invariant during self-propelled swimming.  Then when the amplitude of motion in increased, the Strouhal number will increase and, consequently, the strength of the wake vortices will increase.  In this case, the change in pressure across the pitching airfoil will scale with the strength of the vortices, that is, $\Delta P \propto \rho \Gamma^2/c^2$.  The circulation of the shed vortices will scale as $\Gamma_w \propto fAc$ by invoking unsteady thin airfoil theory.  This characteristic pressure will act on the projected frontal area.  Since this area varies in time, it should scale with the amplitude of motion and the unit span length, $s$, that is, $D_{form} \propto \rho s f^2 A^3$.  Now, if this is non-dimensionalized by the added mass forces (Eq. \ref{coeffs}) and added as a drag term to a Garrick-inspired scaling relation (Eq. \ref{CT_Garrick}) then the following thrust coefficient scaling relation is determined,
\begin{align} \label{CTstar}
C_T(k,\As) &= c_1 \, - c_2 \, w(k) - c_3\, \As 
\end{align}

This modified Garrick relation indicates that the thrust should collapse to a \textit{flat plane} in three-dimensions when it is graphed against $w(k)$ and $\As$.  Indeed, Figures \ref{thrust}c and \ref{thrust}d show excellent collapse of the data to a plane, which can be accurately assessed by viewing the plane of data ``edge-on" (Figure \ref{thrust}d).  In fact, the data collapse has been improved over (\ref{CT_Garrick}) by including the mild form drag modification to Garrick's relation.  Using this scaling relation is a significant deviation from the scaling proposed in \cite{Dewey_2013} and \cite{Quinn2014} and a mild deviation from the analytical solution of \cite{Garrick1936}.

The thrust scaling relation can become predictive by determining the unknown coefficients that define the plane.  By minimizing the squared residuals the coefficients are determined to be $c_1 = 2.89$, $c_2 = 4.02$ and $c_3 = 0.39$ leading to a maximum error of 6.5\% between the scaling relation (reference plane) and the data. These coefficients are not likely to be universal numbers, but may depend upon the shape or thickness of the airfoil.  However, since the data is defined by a flat plane then only \textit{three} simulations would need to be run in order to tune the coefficients such that (\ref{CTstar}) can predict the thrust for a wide range of kinematic parameters.
 
\begin{figure}
\vspace{0mm}
	\begin{center}
	\centerline{ 		{
		\includegraphics[width=0.99\textwidth]{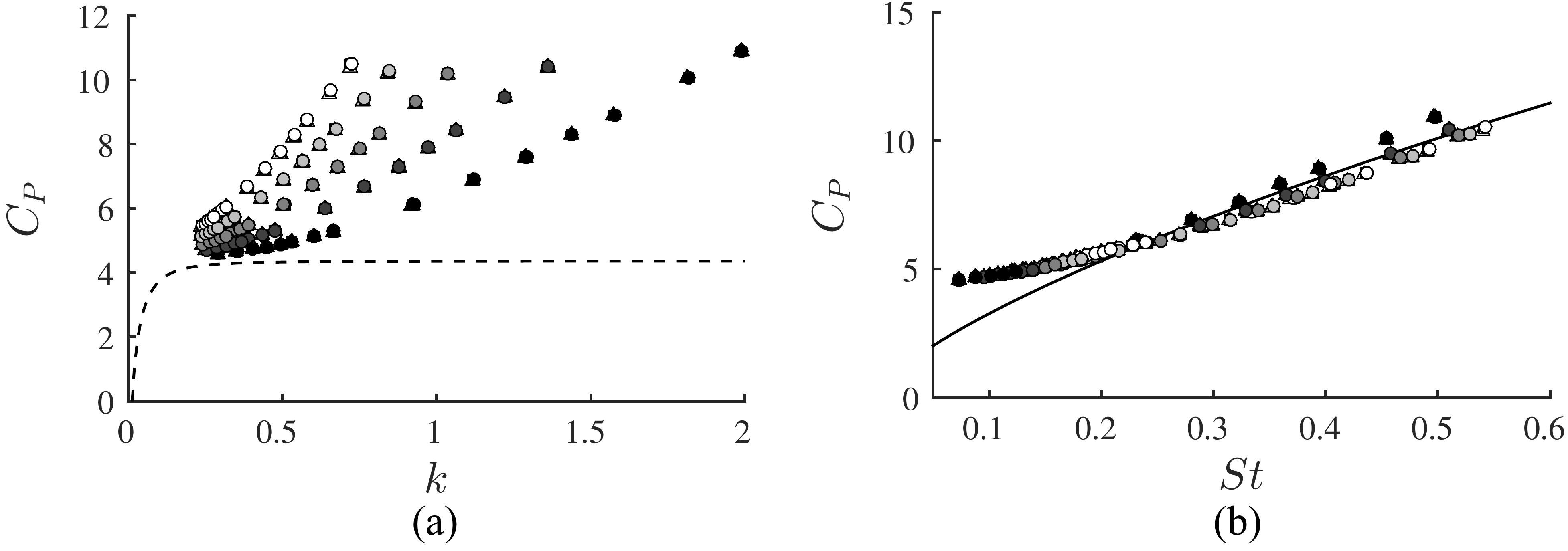}}
			}
	\end{center}
	\caption{(a) Power coefficient as a function of the reduced frequency.  The dashed black line is Garrick's theory.  (b) Power coefficient as a function of the Strouhal number.  The solid black line the scaling relation proposed by \cite[]{Quinn2014}, that is, the best fit line of the following form, $C_P = a\, St^{0.7}$.}
	\label{power}
\end{figure}

Now, we can consider the time-averaged power consumption, which for the three orders of magnitude variation in the frequency, ranges over eight orders of magnitude.  Figure \ref{power}a shows the power coefficient, defined in equation (\ref{coeffs}), as a function of the reduced frequency.  The power coefficient is predicted to be solely a function of $k$ by Garrick's theory (dashed black line in Figure \ref{power}a), however in contrast to the thrust coefficient, the power coefficient data from the self-propelled simulations does not follow the trend from the theory.  This was noted previously for \textit{fixed velocity} experiments and simulations \cite[]{Quinn2014} and it suggests that a small modification of Garrick's theory will not suffice as a scaling relation for the power consumption.

Previously, \cite{Quinn2014} proposed that the power consumption must include information about the large-amplitude separating shear layer at the trailing-edge with a power scaling following $P_{sep} \propto \rho S_p f^3 A^3$ based on the unsteady dynamic pressure from the lateral velocity scale.  They further suggested that the total power consumption of a swimmer would transition from a Garrick power scaling to this separating shear layer scaling as the amplitude of motion became large.  They accounted for this transition by proposing that the power coefficient scaled as $C_P^{dyn}\propto St^{2+\alpha}$, where $\alpha = 0.7$ was empirically determined.  The power coefficient for this scaling relation was defined with the dynamic pressure, that is, $C_P^{dyn} \equiv \overline{P}/(1/2\; \rho S_p U^3)$.  We can reformulate this relation in terms of the power coefficient normalized by the added mass power,
\begin{align}
C_P^{dyn}\left(\frac{1}{2\, St^2}\right) =  \frac{\overline{P}}{\rho S_p f^2 A^2 U} = C_P
\end{align}

The Garrick power coefficient is shown in Figure \ref{power}b as a function of $St$.  A best fit line following the relation $C_P = a \, St^{0.7}$ is determined by minimizing the squared residuals.  Now, there are two observations that can be made.  First, the previously determined scaling relation reasonably captures the power consumption at high $St$, however, there is up to a 50\% error in the relation at low $St$.  In self-propelled swimming  the low $St$ regime corresponds to the lowest $Li$ and consequently the fastest swimming speeds making it an important regime.  Also, the less than linear relation predicted by the previous relation does not capture the nearly linear or greater than linear trend in the data.  The second observation is that at high $St$ the data has a clear stratification based on the amplitude of motion.  This implies that the previous scaling relation does not capture all of the physics that determine the power consumption for an unsteady swimmer.  
\begin{figure}
\vspace{0mm}
	\begin{center}
	\centerline{ 		{
		\includegraphics[width=0.99\textwidth]{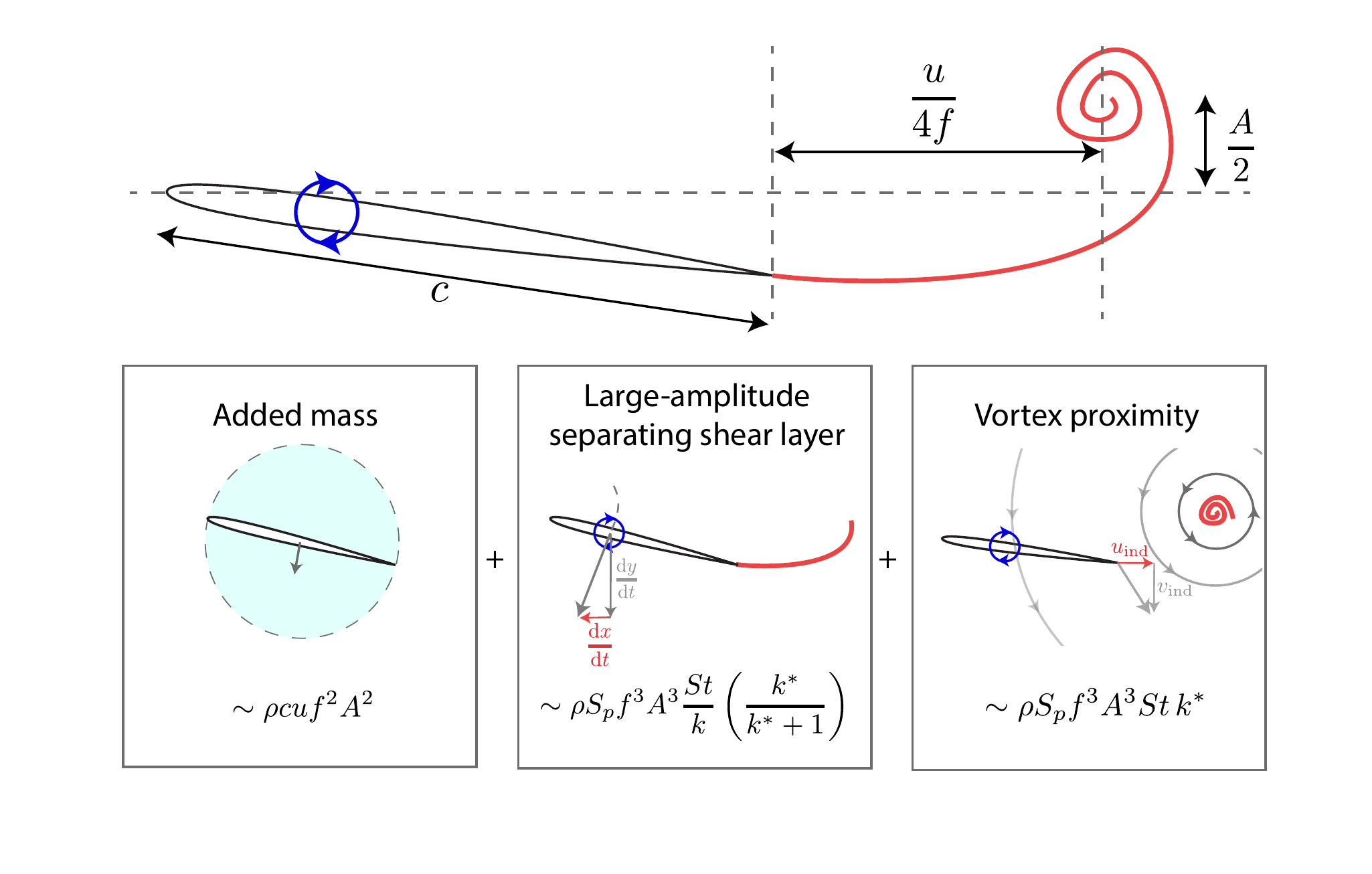}}
			}
	\end{center}
	\caption{Schematic showing the components of the proposed novel power scaling relation.}
	\label{scalingschem}
\end{figure}

To overcome these limitations we now propose that the total power consumption will instead be a \textit{linear combination} of the added mass power scaling, a power scaling from the large-amplitude separating shear layer and a power scaling from the proximity of the trailing-edge vortex (Figure \ref{scalingschem}) as,
\begin{align} \label{lin_comb}
\overline{P} = c_4\, \overline{P}_{add} + c_5\, \overline{P}_{sep} + c_6 \, \overline{P}_{prox}.
\end{align}

For the Garrick portion of the power scaling the exact relation could be used, however, only the added mass contribution to the power is hypothesized to be important since the circulatory component affects the power only at very low reduced frequencies outside of the range of the data in the current study (see Figure \ref{power}a).  This added mass power from Garrick's theory scales as,
\begin{align}
\overline{P}_{add} \propto \rho S_p f^2 A^2 U.
\end{align}

To determine a power consumption scaling due to the large-amplitude separating shear layer consider that in a large-amplitude pitching motion there is an $x$-velocity to the motion of the airfoil.  This velocity component disappears in the small-amplitude limit, but during finite amplitude motion it adds an additional velocity component acting on the bound vorticity of the airfoil that scales as $dx/dt \propto c \dot{\theta}\, sin\, \theta$.  This in turn leads to an additional contribution to the vortex force, or generalized Kutta-Joukowski force, acting on the airfoil.  The lift from this additional term would scale as, $L_{sep} \propto \rho s \Gamma_b\, dx/dt$, where $\Gamma_b$ is the bound circulation of the airfoil.  Following \cite{McCune1990}, the bound circulation can be decomposed into the quasi-steady circulation, $\Gamma_0$, and the additional circulation, $\Gamma_1$ as $\Gamma_b = \Gamma_0 + \Gamma_1$.  The quasi-steady bound circulation is the circulation that would be present from the airfoil motion alone \textit{without} the presence of the wake while the additional circulation arises to maintain a no-flux boundary condition on the airfoil surface and the Kutta condition at the trailing-edge in the presence of a wake.  Now, the quasi-steady circulation scales as $\Gamma_0 \propto c^2 \dot{\theta}$ (neglecting the quasi-static $sin\, \theta$ contribution), so that 
\begin{align} \label{P_sep}
P_{sep} \propto \rho s c \, (c \dot{\theta})^3 \, sin\, \theta + \rho s \Gamma_1 \, (c \dot{\theta})^2\, sin\, \theta
\end{align}

Additionally, it has been recognized \cite[]{Wang2013,Liu2014} that a phase shift in the circulatory forces arises due to nonlinearities introduced by the large-amplitude deforming wake such that some circulatory power contributions may no longer be orthogonal in the time-average.  Consequently, when the time-average of relation (\ref{P_sep}) is taken the second term can be retained, however, the first term will remain orthogonal since by definition it is unaffected by the nonlinearities of the wake.  Now, the time-averaged power from the large-amplitude separating shear layer scales as,
\begin{align} \label{P_sep_TA}
\overline{P}_{sep} \propto \rho s \Gamma_1 \, (c \dot{\theta})^2\, sin\, \theta.
\end{align}

\noindent It becomes evident that this term only exists when there is \textit{large-amplitude motion} where $dx/dt$ is finite and when there is a nonlinear \textit{separating shear layer} or wake present such that orthogonality does not eliminate the additional circulation term in the time-average. This effect is what justifies the use of ``\textit{large-amplitude separating shear layer}" when describing this explicitly nonlinear term.

To complete the scaling, a relation for the additional circulation is needed.  To first order, the additional bound circulation can be modeled as a point vortex located at $x_0$ along the chord.  To estimate the location $x_0$, consider the quasi-steady circulation represented as a point vortex of strength $\Gamma_0$.  During pitching motions about the leading-edge the trailing-edge cross-stream velocity scales as $c\dot{\theta}$.  The Kutta condition is enforced when the induced velocity from the quasi-steady bound vortex counteracts the cross-stream velocity, that is, $c\dot{\theta} = \Gamma_0/(2\pi x_0)$.  Since $\Gamma_0 = \pi\, c^2 \dot{\theta}/2$, the bound vortex location is determined to be $x/c = 3/4$.  By using this location as an estimate of the location of a point vortex with strength $\Gamma_1$ (additional circulation) the Kutta condition at the trailing edge can be preserved by matching the induced velocity from $\Gamma_1$ with the induced velocity from the trailing-edge vortex (TEV).  Following Figure \ref{scalingschem}, consider the formation of a TEV positioned a distance of $A/2$ and $U/4f$ from the trailing edge in the cross-stream and downstream directions, respectively.  The strength of the TEV is proportional to $\Gamma_w = \Gamma_0 + \Gamma_1$ from Kelvin's condition.  This TEV induces a velocity at the trailing edge with both a $u$ and $v$ component.  For now, consider that the $v$ component must be canceled at the trailing-edge by the additional bound vortex $\Gamma_1$.  By following these arguments a scaling relation for the additional circulation is determined,
\begin{align}\label{Gamma_1}
\Gamma_1 \propto c^2 \dot{\theta} \left( \frac{\ks}{\ks + 1}\right), \quad \quad \quad \mbox{where: } \ks\equiv \frac{k}{1 + 4\, St^2}
\end{align}

\noindent Since the factor of $\ks$ ranges from $0.17 \leq \ks/(\ks+ 1) \leq 0.5$ for the data from the current study, the additional circulation can, to a reasonable approximation, be considered as a perturbation to the quasi-steady circulation, $\Gamma_0 \propto c^2 \dot{\theta}$.  Now by substituting relation (\ref{Gamma_1}) into (\ref{P_sep_TA}) the following power relation is obtained,
\begin{align}
\overline{P}_{sep}\propto \rho S_p \, f^3 A^3 \frac{St}{k} \left( \frac{\ks}{\ks + 1}\right) .
\end{align}

The second explicitly nonlinear correction to the power is due to the proximity of the TEV.  In unsteady linear theory the wake is considered to be non-deforming and planar.  As such it only induces an upwash or downwash over the thin airfoil, that is, a $v$ component of velocity.  In a fully nonlinear numerical solution and in real flows, the vortex wake is deformed and does not lie along a plane except in some special cases.  Consequently, the wake vortices induce an additional $u$ velocity, which adds an additional lift and power contribution through the vortex force.  In this way, the lift from the proximity of the TEV would scale as,
\begin{align}
L_{prox} \propto \rho s u_{\mathrm{ind}} (\Gamma_0 + \Gamma_1)
\end{align}

\noindent By considering the same schematic used in the additional circulation scaling (Figure \ref{scalingschem}), the $u$ velocity induced at the trailing-edge by the TEV would scale as, $u_{\mathrm{ind}} \propto c^2 \dot{\theta}\, f\, St / \left[U \left( 1 + 4\, St^2\right)\right]$.  In contrast to the large-amplitude separating shear layer correction, the TEV induced velocity does not have a $\mathrm{sin}\, \theta$ and thereby the quasi-steady term is retained in the time-average.  Then to a reasonable approximation the additional circulation can be neglected since it scales with the quasi-steady circulation and its magnitude is a fraction of the magnitude of the quasi-steady circulation.  Consequently, the power scaling from the proximity of the trailing-edge vortex is
\begin{align}
\overline{P}_{prox} \propto \rho S_p f^3 A^3 St\, \ks,
\end{align}

\noindent and the total power consumption scaling relation is
\begin{align}
\overline{P} = c_4\, \rho S_p f^2 A^2 U + c_5\,\rho S_p \, f^3 A^3 \frac{St}{k}\left( \frac{\ks}{\ks + 1}\right)  + c_6\, \rho S_p f^3 A^3 St\, \ks.
\end{align}

\noindent When this power is nondimensionalized by the added mass power, the following scaling relation is determined for the coefficient of power:
\begin{align} \label{pow_scale}
C_P\left(k,St \right) &= c_4 + c_5\, \frac{St^2}{k}\left(\frac{\ks}{\ks + 1}\right) + c_6\, St^2 \ks.
\end{align}

\noindent It is evident that this relation can be rewritten as,
\begin{align} \label{pow_scale_plane}
C_P\left(k,St \right) &= c_4 + c_5\, \phi_1 + c_6\, \phi_2  \\
  \text{where: } \quad \phi_1 &= \frac{St^2}{k}\left(\frac{\ks}{\ks + 1}\right), \quad \phi_2 = St^2 \ks. \nonumber
\end{align}

\begin{figure}
\vspace{0mm}
	\begin{center}
	\centerline{ 		{
		\includegraphics[width=0.99\textwidth]{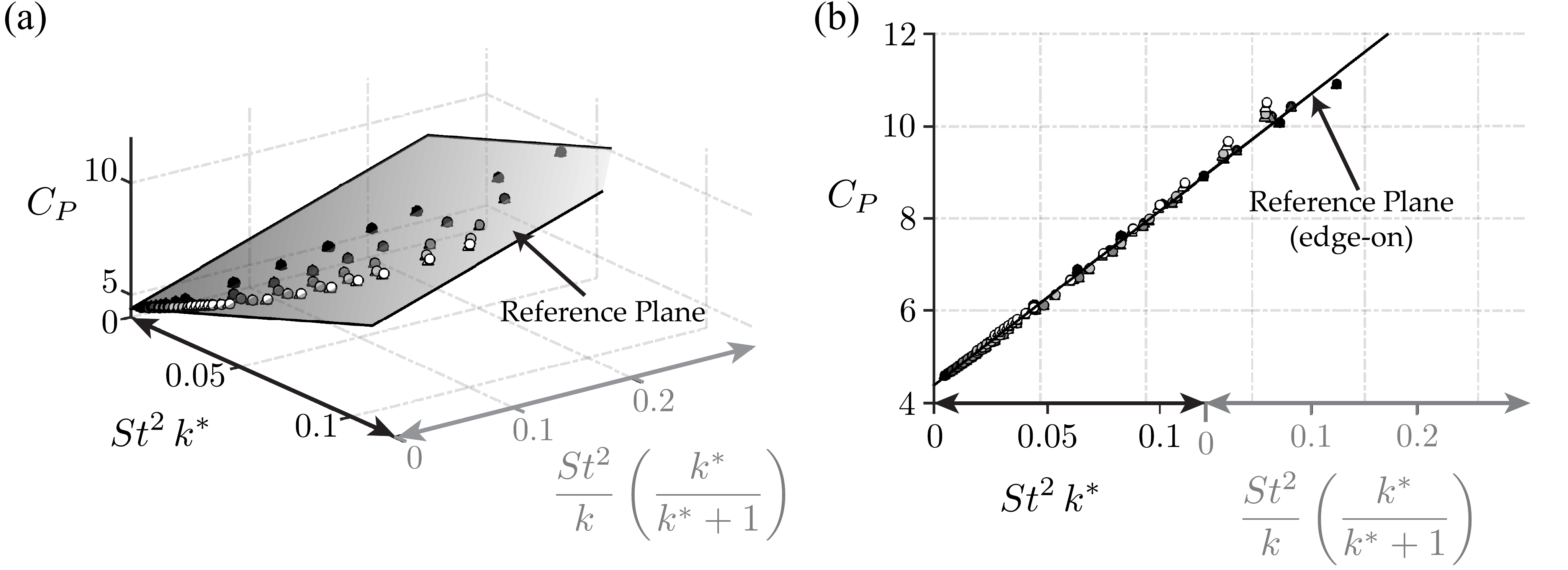}}
			}
	\end{center}
	\caption{(a) Three-dimensional graph of the power coefficient as a function of the large amplitude separating shear layer and vortex proximity scaling terms. (b) Three-dimensional power coefficient graph oriented edge-on with a reference plane.  (c) Power coefficient as a function of the right-hand side of (\ref{pow_scale}).  (d) Classic power coefficient nondimensionalized by the dynamic pressure.  In both (c) and (d) the solid black line is the scaling relation (\ref{pow_scale}).}
	\label{power_scale}
\end{figure}

\noindent If the power scaling relation is accurate, then equation (\ref{pow_scale_plane}) states that the data should collapse to a \textit{flat plane} when $C_P$ is graphed as a function of $\phi_1$ and $\phi_2$.  Figure \ref{power_scale}a shows $C_P$ as a function of $\phi_1$ and $\phi_2$ in a three-dimensional graph.  Indeed, by rotating the orientation of the data about the $C_P$-axis (Figure \ref{power_scale}b) such that the view is edge-on with the reference plane it then becomes clear that there is an excellent collapse of the $C_P$ data to a plane.   This indicates that the power scaling relation accounts for most of the missing nonlinear physics not present in linear unsteady airfoil theory.  In fact, the missing nonlinear terms are essentially the two parts of the explicitly nonlinear term described in \cite{McCune1990} and \cite{McCune1993}. 

A quantitatively predictive relation can be obtained by further determining the unknown coefficients in (\ref{pow_scale}).  Importantly, since the power data collapses to a plane then only \textit{three} simulations or experiments would need to be run in order to make (\ref{pow_scale}) predictive for a wide range of kinematic parameters.  By minimizing the squared residuals, the coefficients are determined to be $c_4 = 4.38$, $c_5 = 37.9$ and $c_6 = 17.8$.  Then the largest error between the scaling relation (reference plane) and the data is 5.6\% as opposed to 50\% with the previous scaling relation of \cite{Quinn2014}, showing an order-of-magnitude improvement in the accuracy of the scaling.  

Finally, it is expected that the thrust and power scaling relations are valid for both self-propelled swimming and fixed-velocity swimming, although the coefficients may differ.  In fixed-velocity swimming it is evident from the scaling relations that the thrust and power coefficients and subsequently the efficiency are only dependent upon the reduced frequency, $k$, and the Strouhal number, $St$ (note that $\As = St/k$).

\section{Self-Propelled Swimming Scaling Relations}
The major contributions of this study are the thrust and power scaling relations during self-propelled swimming. These relations are now algebraically extended to determine the mean swimming speed and cost of transport during self-propelled swimming, which are output parameters of interest in the design of bio-inspired devices.  First, the relations to predict the mean swimming speed, reduced frequency and Strouhal number will be determined.  Then the scaling relations for the energetic metrics of efficiency, economy and cost of transport will be developed.

\subsection{Mean Swimming Speed, Reduced Frequency and Strouhal Number}
Once a steady-state of the cycle-averaged swimming speed has been reached the time-averaged thrust force must balance the time-averaged drag force, that is $\overline{T} = \overline{D}$ \cite[]{Lighthill1960}.  Biological swimmers that operate at $Re > \mathcal{O}(10^4)$ follow a $U^2$ drag law as shown in eq. (\ref{eq:draglaw}) \cite[]{Gazzola2014}.  Now, by setting the time-averaged thrust and drag scaling relations equal to each other and introducing the appropriate constants of proportionality we arrive at a prediction for the cruising speed,
\begin{align} \label{U_scale}
\overline{U}_p &= fA \sqrt{2 \frac{C_T(k,\As)}{Li}}
\end{align}

\noindent It is now clear that $\overline{U} \propto f A$.  As a propulsor scales to a larger size when $\As$ is fixed, $A$ will increase and so to will the swimming speed if $f$ remains fixed as well.  However, the mean speed scaling with the amplitude is unclear if $\As$ varies.  Also the scaling of the mean speed with $Li$ is not so straightforward as discussed below.  

The relation eq. (\ref{U_scale}) depends on some parameters known \textit{a priori}, that is $f, \,A,\, Li$.  Still $k$ is not known \textit{a priori} since it depends upon the mean speed.  However, by substituting the mean speed on the left hand side of eq. (\ref{U_scale}) into the reduced frequency relation (\ref{nonDfreq}) and rearranging, the following nonlinear equation for the predicted reduced frequency, $k_p$, can be solved by using any standard iterative method,
\begin{align} \label{k}
k_p^2\left[ c_1 - c_2\, w(k_p) - c_3\, \As \right] -  \frac{Li}{2 {\As}^2}  = 0
\end{align}
\begin{figure}
\vspace{0mm}
	\begin{center}
	\centerline{ 		{
		\includegraphics[width=0.9\textwidth]{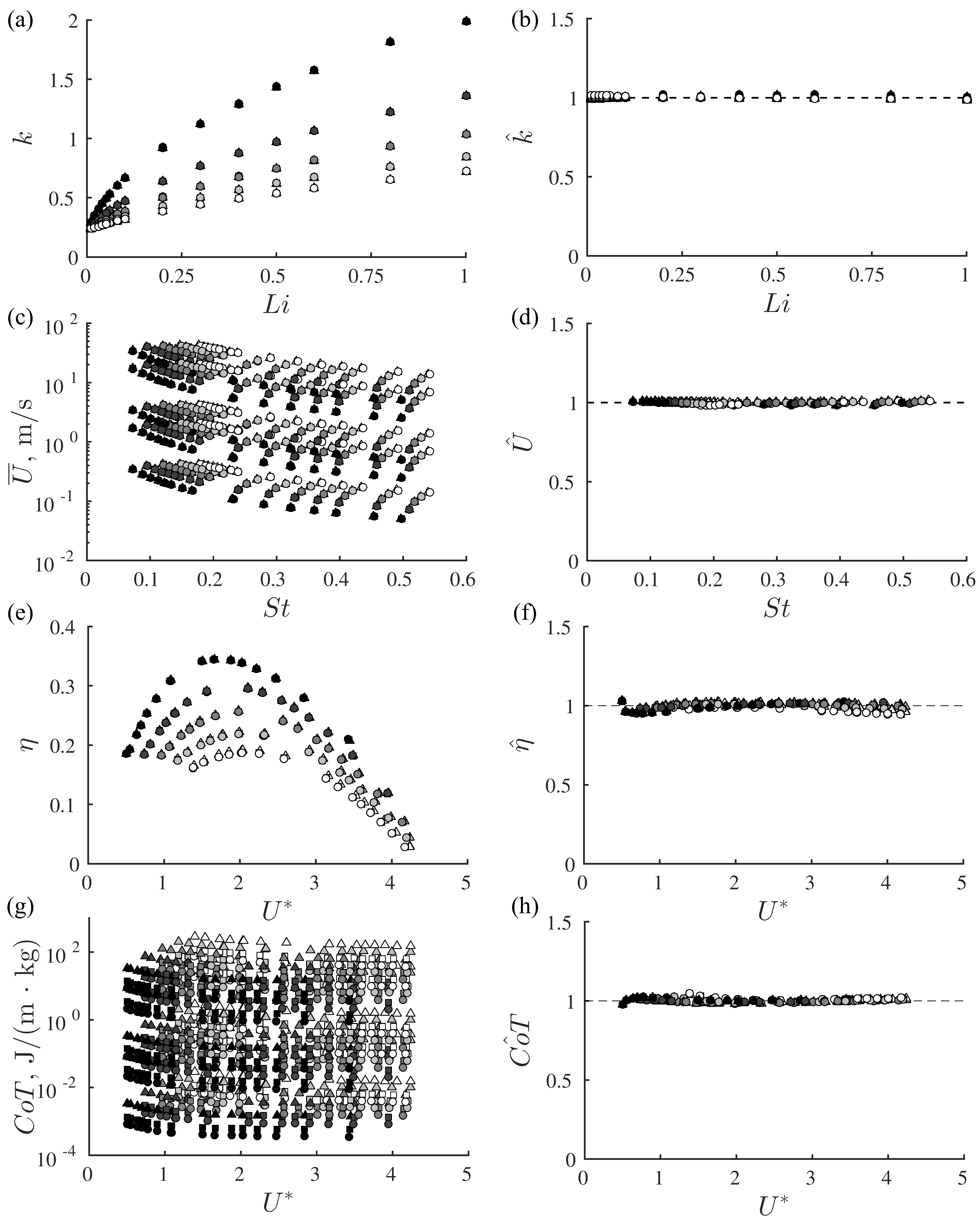}}
			}
	\end{center}
	\caption{(a) Reduced frequency as a function of Lighthill number.  (b) Normalized reduced frequency as a function of Lighthill number. (c) Mean speed as a function of the Strouhal number.  (d) Normalized mean speed as a function of Strouhal number. (e) Propulsive efficiency as a function of the non-dimensional stride length. (f) Normalized efficiency as a function of the non-dimensional stride length. (g) Cost of transport and a function of the non-dimensional stride length. (h) Normalized cost of transport as a function of the non-dimensional stride length.}
	\label{selfpropscale}
\end{figure}

\noindent This relation shows that $k$ is only a function of $\As$ and $Li$, and it does not vary with $f$ or $m^*$ as observed in Figure \ref{selfpropscale}a.  For high reduced frequencies ($k > 1$), $k \propto Li^{1/2}$ and $k \propto {\As}^{-1} [a - b\, \As]^{-1/2}$, where $a$ and $b$ are constants.  Once the predicted reduced frequency is determined, the predicted Strouhal number, $St_p$, and the predicted modified reduced frequency, $\ks_p$ can be directly determined.
\begin{align}
St_p &= k_p\, \As   \label{St}\\
\ks_p &= \frac{k_p}{1 + 4\, St_p^2}  \label{k_star}
\end{align} 

\noindent Again, $St$ and $\ks$ are only functions of $\As$ and $Li$.  Now $k_p$, $St_p$, $\ks_p$ and $\overline{U}_p$ can be determined \textit{a priori} by solving the nonlinear equation (\ref{k}) and using (\ref{St}), (\ref{k_star}) and (\ref{U_scale}).   As observed in Section \ref{sc:results}, varying the frequency will not change the reduced frequency nor the Strouhal number during self-propelled swimming.

The actual reduced frequency from the self-propelled simulations normalized by the predicted reduced frequency, that is $\hat{k} = k/k_p$, is shown in Figure \ref{selfpropscale}b.  It can be seen that there is excellent agreement between the actual and the predicted reduced frequency since $\hat{k} \approx 1$.   The Strouhal number measured from the simulations compared to the predicted Strouhal number has similar agreement since $St = k\, A^*$.

Figure \ref{selfpropscale}c shows the mean speed as a function of the Strouhal number.  The swimming speeds are observed to be spread over four orders of magnitude with the highest swimming speeds occurring in the low $St$ range of $0.075 < St < 0.2$.  Figure \ref{selfpropscale}d shows an excellent collapse of the data with the predicted mean swimming speed being within 2\% of the full-scale value of the actual mean velocity.  Again, this mean speed prediction is calculated \textit{a priori} by using physics-based scaling relations.  

\subsection{Energetics}
It is equally important to be able to predict the amount of energy consumed by an unsteady propulsor during locomotion.  One energetic metric, the propulsive efficiency, is simply the ratio of the thrust and power coefficient, that is,
\begin{align}
\eta_p = \frac{C_T(k,\As)}{C_P(k,St)} = \frac{c_1 \, - c_2 \, w(k) - c_3\, \As  }{c_4 + c_5\, \frac{St^2}{k}\left(\frac{\ks}{\ks + 1}\right) + c_6\, St^2 \ks}
\end{align}

\noindent  This relation indicates that $\eta$ is only a function of $\As$ and $Li$, and it does not depend upon $f$ or $m^*$.  This is reflected in the data presented in Figure \ref{selfpropscale}e with the exception that there is some mild variation in $\eta$ and $U^*$ with $m^*$ for the lowest amplitudes and highest non-dimensional stride lengths.  For these cases, increasing $m^*$ leads to a small increase in $\eta$ and a small decrease in $U^*$.  By comparing the predicted efficiency from this scaling relation to the actual efficiency from the computations, $\hat{\eta} = \eta/\eta_p$, Figure \ref{selfpropscale}f shows that the prediction agrees very well with the actual efficiency.  In fact, the scaling relation can predict the efficiency to within 5\% of its full-scale value.  


After substituting the scaling relations into the economy and cost of transport relations, the prediction for $\xi$ and $CoT$ become
\begin{align}
\xi_p &= \frac{1}{\rho S_p f^2A^2} \frac{1}{\left[c_4 + c_5\, \frac{St^2}{k}\left(\frac{\ks}{\ks + 1}\right) + c_6\, St^2 \ks\right]} \\
CoT_p &= \frac{ \rho S_p f^2A^2}{m}\left[c_4 + c_5\, \frac{St^2}{k}\left(\frac{\ks}{\ks + 1}\right) + c_6\, St^2 \ks\right]
\end{align}

\noindent Figure \ref{selfpropscale}g and \ref{selfpropscale}h show that the $CoT$ relation can accurately predict the cost of transport that spans six orders of magnitude to within 5\% of its full scale value.

\subsection{Scaling Relation Summary and Discussion} \label{sec:summary}
Throughout this study a $U^2$ drag law imposed a drag force on a self-propelled pitching airfoil and this drag relation was used in the development of self-propelled scaling relations.  However, a low $Re$ Blasius drag law that follows a $U^{3/2}$ scaling such as,
\begin{align}
\overline{D} = C_D S_w \left(\frac{\mu \rho}{L}\right)^{1/2} U^{3/2}
\end{align}
\begin{table}
   \begin{center}
   	\begin{threeparttable}[b]
	{
	\renewcommand{\arraystretch}{1.5}
	\begin{tabular}{c|ccc} 
  	 Variables & \hspace{10pt} $10 \leq Re \leq 10^4$ \hspace{10pt} && $Re > 10^4$  \vspace{-6pt}  \\ \hline 
	 & &  \vspace{-21pt} \\
         $\overline{T}=$& \multicolumn{3}{c}{$\rho \, C_T S_p f^2 A^2$}     \\
         $C_T =$& \multicolumn{3}{c}{$c_1 - c_2\, w(k) - c_3\,\As $}     \\    
         $\overline{P} =$& \multicolumn{3}{c}{$\rho \, C_P S_p f^2 A^2 \overline{U}$}  \\ [1ex]
         $C_P =$& \multicolumn{3}{c}{$c_4 + c_5\, \frac{St^2}{k}\left(\frac{\ks}{\ks + 1}\right) + c_6\, St^2 \ks$}     \\ 
	 $\eta =$&  \multicolumn{3}{c}{$\displaystyle \frac{c_1 \, - c_2 \, w(k) - c_3\, \As  }{c_4 + c_5\, \frac{St^2}{k}\left(\frac{\ks}{\ks + 1}\right) + c_6\, St^2 \ks}$}   \\[3ex]
	$\overline{U} =$&   $\displaystyle \left(fA\right)^{4/3} \left(\frac{C_T}{Li}\right)^{2/3}\left(\frac{L}{\nu}\right)^{1/3}$ 	&\hspace{30pt} & $\displaystyle fA \left(2 \frac{C_T}{Li} \right)^{1/2}$ \\[2ex]
	$k\tnote{\footnotemark}$ :& $\displaystyle k\, C_T^{2/3} = {\As}^{-1} Sw^{-1/3} Li^{2/3}$ & &$\displaystyle k^2\, C_T = \frac{Li}{2} {\As}^{-2}$  \\[2ex]
	$St\tnote{\textdagger}$ :& $\displaystyle St\, C_T^{2/3} = Sw^{-1/3} Li^{2/3}$ & & $\displaystyle St^2\, C_T = \frac{Li}{2}$  \\ [1ex]
	$\ks =$& \multicolumn{3}{c}{$\displaystyle \frac{k}{1 + 4\, St^2} $}  \\  [3ex]
	$\xi =$&  \multicolumn{3}{c}{$\displaystyle \frac{1}{\rho S_p f^2A^2} \frac{1}{\left[c_4 + c_5\, \frac{St^2}{k}\left(\frac{\ks}{\ks + 1}\right) + c_6\, St^2 \ks\right]}$} \\ [5ex]
	$CoT =$&  \multicolumn{3}{c}{$\displaystyle\frac{ \rho S_p f^2A^2}{m}\left[c_4 + c_5\, \frac{St^2}{k}\left(\frac{\ks}{\ks + 1}\right) + c_6\, St^2 \ks\right]$} \\ [2ex]
	coefficients :&$c_1 = 2.89$&\hspace{-17pt}$c_2 = 4.02$&$c_3 = 0.39$     \\
	&$c_4 =4.38$&\hspace{-17pt}$c_5 = 37.9$&$c_6 = 17.8$    \\
	\end{tabular} 
	}
	\begin{tablenotes}
    		\item[\textdagger] {\footnotesize These equations must be solved with an iterative method.}
 	\end{tablenotes}
      \end{threeparttable}
   \end{center} 	
  	\caption{Summary of scaling relations.  Both high and low Reynolds number scalings are presented.  The swimming number is $Sw = f A L / \nu$.}
 	\label{tab:scalings}
\end{table}

\noindent may be used for swimmers that operate in the Reynolds number range of $\mathcal{O}(10) \leq Re \leq \mathcal{O}(10^4)$ \cite[]{Gazzola2014}.  To account for this drag regime, the scaling relations developed in this study are reformulated for a Blasius drag law and both drag law cases are summarized in Table \ref{tab:scalings}.  Some of these relations use the swimming number first defined in \cite{Gazzola2014}.  Here it is slightly different and is $Sw = f A L / \nu$.  One major conclusion of this study is that in the high $Re$ regime, the \textit{nondimensional performance} of a self-propelled two-dimensional pitching airfoil is completely defined by two nondimensional variables, namely, the Lighthill number, $Li$, and the amplitude-to-chord ratio, $\As$.  Similarly, in the low $Re$ regime the \textit{nondimensional performance} is completely defined by $Li$, $\As$ and the swimming number, $Sw$.  Although the high $Re$ scaling relations would benefit from being cast in terms of $Li$ and $A^*$ explicitly, this is unfortunately impossible given the nonlinear relationship between $k$, $St$ and $Li$, $A^*$.

The mean thrust and power, their coefficients and the propulsive efficiency are all independent of the drag regime.  These scaling relations form the basis for the subsequent self-propelled relations.  The relations developed in the current study differ from previous work \cite[]{Garrick1936,Dewey2011,Eloy2012a,Quinn2014,Gazzola2014} by considering that the thrust scales with both added mass and circulatory forces and by considering that the power scales with a combination of the power incurred by added mass forces and vortex forces from large amplitude motion in the presence of a separating shear layer and from the induced velocity of the trailing-edge vortex.

The scaling relation for the mean speed, reduced frequency and Strouhal number are all affected by the drag regime.  The mean speed relations are seen to be similar to the study of \cite{Gazzola2014} except that we consider the thrust coefficient and Lighthill number of the swimmer.  If these quantities are constant then the relations from \cite{Gazzola2014} are recovered, however, both of these numbers vary across species \cite[]{Eloy2012a} and are design parameters for devices.  Consequently, the relations in the current study are in fact a generalization of the relations presented in \cite{Gazzola2014}.

One surprising revelation is that the swimming economy and cost of transport scaling relations are independent of the drag regime of the swimmer.  This leads to a new non-dimensionalization of $CoT$ as,
\begin{align}
CoT^* &\equiv \frac{CoT\, m}{\rho S_p f^2A^2} \\
\mbox{and} \quad CoT^* &= C_P = \left[c_4 + c_5\, \frac{St^2}{k}\left(\frac{\ks}{\ks + 1}\right) + c_6\, St^2 \ks\right]
\end{align}

\noindent In the case of this data, the added mass power accounts for nearly six orders of magnitude of variation, while the power coefficient varies only over a factor of three (see Figure \ref{power_scale}a).  In biology, the cost of transport varies over three orders of magnitude \cite[]{Videler1993} and the Strouhal number varies by about a factor of two ($0.2 \leq St \leq 0.4$, \cite{Taylor_2003}).  

\subsection{Comparison with Biology}  
\begin{figure}
\vspace{0mm}
	\begin{center}
	\centerline{ 		{
		\includegraphics[width=0.99\textwidth]{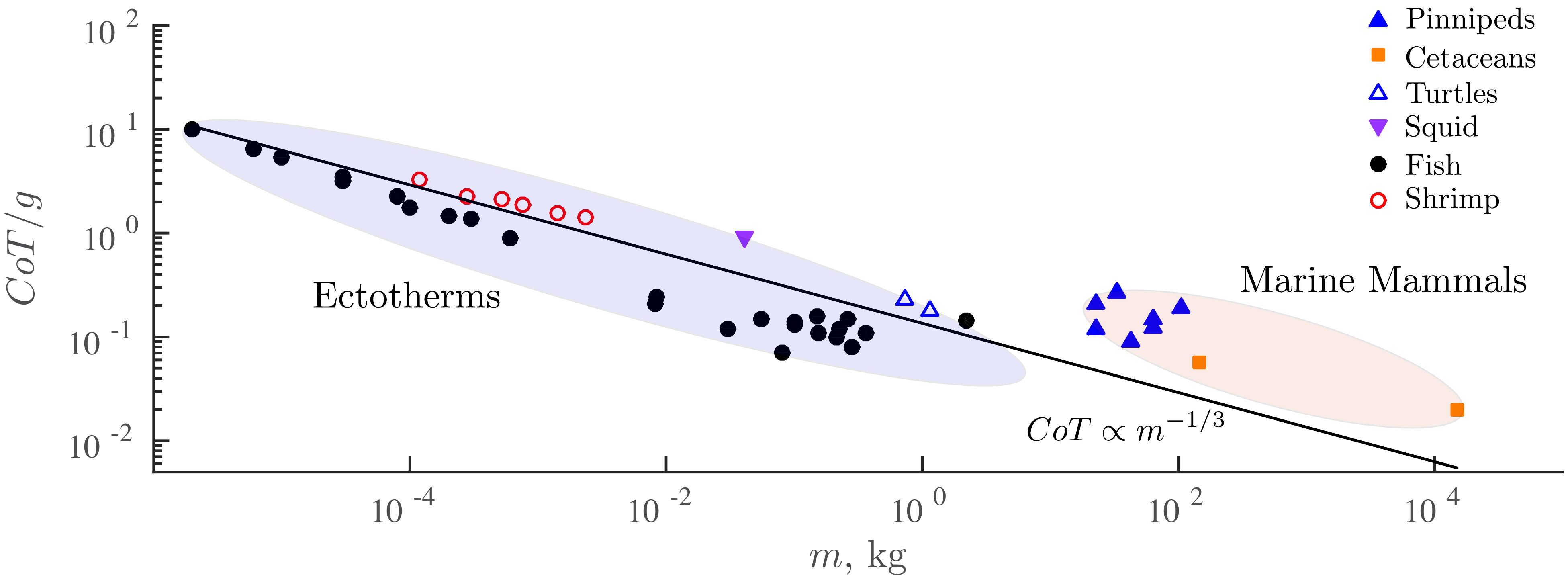}}
			}
	\end{center}
	\caption{Log-log graph of the net cost of transport normalized by the acceleration due to gravity (total cost of transport minus the cost of transport due to the resting metabolic rate of an animal) as a function of mass.  There are 34 ectothermic individuals including fish, shrimp, squid and turtles, and 11 marine mammals including seals, sea lions, dolphins and whales.  The ectotherm data are from \cite[]{Videler1990,Kaufmann1990,Videler1993,Dewar1994} and the marine mammal data are from \cite[]{Videler1990,Videler1993,Williams1999}.}
	\label{BioData}
\end{figure}

We have shown that the cost of transport of a self-propelled pitching airfoil will scale predominately as $CoT \propto \rho S_p f^2 A^2/m$, which means that the orders of magnitude variation in $CoT$ is scaling with the power input from added mass forces.  If a virtual body combined with a self-propelled pitching airfoil is a reasonable model of unsteady animal locomotion then a similar $CoT$ scaling would be expected in biology.  To compare this added mass power-based scaling with biological data let's consider how the $CoT$ relation scales with the length of a swimmer, $L$.  The parameters in the relation scale as $S_p \propto L^2$, $m \propto L^3$,  $A \propto L$ \cite[]{Bainbridge1957}, and for swimmers $f \propto L^{-1}$ \cite[]{Sato2007,Broell2015}.  By combining these the $CoT \propto L^{-1}$ and $CoT \propto m^{-1/3}$ and the range of a self-propelled device or animal should scale as $\mathcal{R}  \propto L$ and $\mathcal{R}  \propto m^{1/3}$ for a fixed energy density.  Indeed, the net cost of transport, that is the total $CoT$ minus the $CoT$ due to the resting metabolic rate of an animal, has a scaling of $CoT \propto m^{-0.36}$ for 34 ectothermic individuals including fish, shrimp, squid and turtles, and a scaling of $CoT \propto m^{-0.34}$ for 11 marine mammals including seals, sea lions, dolphins and whales (see Figure \ref{BioData}), which are very similar to the scaling of $CoT \propto m^{-1/3}$ proposed here.  The scaling relation for $CoT$ further suggests that variations in the Strouhal number and reduced frequency of swimming animals can have a mild effect on the scaling of $CoT$.  To the best of the authors knowledge, the scaling of $CoT \propto m^{-1/3}$ is the first energetic scaling that aligns with biological data and is based on a mechanistic rationale.

\subsection{Model Limitations and Scaling Relation Extensions}
One potential limitation of the current study is that a drag law was imposed and not solved for as is the case in Navier-Stokes solutions.  However, the choice of the drag law does not alter the thrust coefficient, power coefficient, efficiency, and cost of transport scaling relations.  Choosing a different drag law only changes the range of $k$ and $St$, but the sufficient criteria is that these ranges fall within ranges observed in nature as is the case in the current study. 

The drag law is applied to a virtual body, which acts as the simplest representation of the effect of coupling a body to a propulsor in self-propelled swimming.  This approach does not consider the interaction of body vorticity with propulsor vorticity and as such many effects of this interaction are not considered \cite[]{Zhu2002,Akhtar2007}.  Furthermore, the airfoils in the current study are two-dimensional, rigid, oscillate with pitching motions, and operate in an inviscid environment, yet the physics are still quite complex.  

Importantly, a new scaling relation framework has been developed that can be extended to include many of the neglected physical phenomena.  Recent and ongoing studies are extending this basic propulsion study by incorporating (1) two-dimensional heaving and combined heaving and pitching effects, (2) three-dimensional effects, (3) flexibility effects, and (4) intermittent swimming effects. All of these research directions could benefit from the improved scaling presented in the current study.  First, extending the scaling relations to combined heaving and pitching can follow the same procedure as the current paper by determining where Garrick's linear theory is no longer valid and providing scaling relations for the nonlinear physics that are not accounted for by linear theory.  Second, incorporating three-dimensional effects may occur by considering both the variation in the circulatory forces following a lifting line theory correction \cite[]{Green2008} as well as by considering the alteration of the added mass forces with aspect ratio \cite[]{Dewey_2013}.  Third, flexibility may be accounted for by considering previous scaling relations \cite[]{Dewey_2013}, however, this previous work only examined a range of high reduced frequency where added mass forces dominate.  Finally, the scaling relations from the current study have already been successfully generalized to capture intermittent swimming effects \cite[]{Akoz2017}.   Additionally, future work will focus on extending the scaling relations to consider viscous effects by using both DNS and experiments.

\section{Conclusion}
This study has examined the swimming performance of a two-dimensional pitching airfoil connected to a virtual body.  A scaling relation for thrust was developed by considering both the added mass and circulatory forces, and for power by considering the added mass forces as well as the vortex forces from the large-amplitude separating shear layer at the trailing edge and from the proximity of the trailing-edge vortex.  These scaling relations are combined with scalings for drag in laminar and turbulent flow regimes to develop scaling relations for self-propulsion.  The scaling relations indicate that for $Re > \mathcal{O}(10^4)$ the nondimensional performance of an self-propelled pitching airfoil only depends upon $(Li, \As)$ while for $\mathcal{O}(10) < Re < \mathcal{O}(10^4)$ it depends upon $(Li, \As, Sw)$.  The scaling relations are shown to be able to predict the mean swimming speed, propulsive efficiency and cost of transport to within 2\%, 5\% and 5\% of their full-scale values, respectively, by only using parameters known \textit{a priori} for a NACA 0012 airfoil.  The relations may be used to drastically speed-up the design phase of bio-inspired propulsion systems by offering a direct link between design parameters and the expected $CoT$.  The scaling relations also suggest that the $CoT$ of organisms or vehicles using unsteady propulsion should scale predominantly with the power input from added mass forces.  Consequently, their cost of transport will scale with their mass as $CoT \propto m^{-1/3}$, which is shown to be consistent with existing biological data.  This offers one of the first mechanistic rationales for the scaling of the energetics of self-propelled swimming.  

\section{Acknowledgements}
The authors would like to thank the generous support provided by the Office of Naval Research under Program Director Dr. Robert Brizzolara, MURI Grant No. N00014-14-1-0533. We would like to thank Daniel Floryan, Tyler Van Buren and Alexander Smits for their insightful and thought-provoking discussion on the scaling relations.  We would also like to thank Emre Akoz and John Cimbala for their helpful discussions on the power coefficient.

\bibliographystyle{jfm}
\bibliography{ScalingLaws_Literature}

\end{document}